\documentclass[12pt,preprint]{aastex}
\usepackage{natbib}

\shorttitle{MEGA PI}
\shortauthors{Winters et al.}

\begin{document}

\title{The Solar Neighborhood XXXVIII. Results from the CTIO/SMARTS
  0.9m: Trigonometric Parallaxes for 151 Nearby M Dwarf Systems}

\author{Jennifer G. Winters\altaffilmark{1}$^,$\altaffilmark{7},
  R. Andrew Sevrinsky\altaffilmark{2}$^,$\altaffilmark{7}, Wei-Chun
  Jao\altaffilmark{2}$^,$\altaffilmark{7}, Todd
  J. Henry\altaffilmark{3}$^,$\altaffilmark{7}, Adric
  R. Riedel\altaffilmark{4}$^,$\altaffilmark{7}, John
  P. Subasavage\altaffilmark{5}$^,$\altaffilmark{7}, John
  C. Lurie\altaffilmark{6}$^,$\altaffilmark{7}, Philip
  A. Ianna\altaffilmark{3}$^,$\altaffilmark{7}}

\altaffiltext{1}{Harvard-Smithsonian Center for Astrophysics, Cambridge,
  MA 02138; jennifer.winters@cfa.harvard.edu}

\altaffiltext{2}{Department of Physics and Astronomy, Georgia State
  University, Atlanta, GA 30302-4106; sevrinsky@astro.gsu.edu,
  jao@astro.gsu.edu}

\altaffiltext{3}{RECONS Institute, Chambersburg, Pennsylvania,
  17201; toddhenry28@gmail.com, philianna3@gmail.com}

\altaffiltext{4}{Astronomy Department, California Institute of Technology,
  Pasadena, CA 91125; arr@astro.caltech.edu}

\altaffiltext{5}{US Naval Observatory, Flagstaff Station, 10391 West Observatory
Road, Flagstaff, AZ 86001; jsubasavage@nofs.navy.mil}

\altaffiltext{6}{Astronomy Department, University of Washington, Seattle, WA
  98195; lurie@uw.edu}

\altaffiltext{7}{Visiting Astronomer, Cerro Tololo Inter-American
Observatory.  CTIO is operated by AURA, Inc. under contract to the
National Science Foundation.}

 
\begin{abstract}
\label{abstract}

We present 160 new trigonometric parallaxes for 151 M dwarf systems
from the REsearch Consortium On Nearby Stars (RECONS) group's
long-term astrometry/photometry program at the CTIO/SMARTS 0.9m
telescope.  Most systems (124 or 82\%) are found to lie within 25 pc.
The stars have 119 mas yr$^{-1}$ $\leq$ $\mu$ $\leq$ 828 mas yr$^{-1}$
and 3.85 $\leq$ $(V-K)$ $\leq$ 8.47.  Among these are 58 systems from
the SuperCOSMOS-RECONS (SCR) search, discovered via our proper motion
trawls of the SuperCOSMOS digitized archival photographic plates,
while the remaining stars were suspected via photometric distance
estimates to lie nearby. Sixteen are systems that are newly discovered
via astrometric perturbations to be binaries, many of which are ideal
for accurate mass determinations due to their proximity and orbital
periods on the order of a decade.  A variability analysis of the stars
presented, two-thirds of which are new results, shows six of the stars
to vary by more than 20 mmag.  This effort brings the total number of
parallaxes for M dwarf systems measured by RECONS to nearly 500 and
increases by 26\% the number of southern M dwarf systems with accurate
trigonometric parallaxes placing them within 25 pc.


\end{abstract} 


\section{Introduction}
\label{sec:intro}

Historically, trigonometric parallax ($\pi_{trig}$) measurements for M
dwarfs have been a challenge due to the intrinsic faintness of these
low-mass stars. Thus, early astrometric programs typically targeted
the brightest and earliest of M dwarf spectral types, which were
usually also the nearest. The first publication of parallaxes for M
dwarf primaries reported results for stars with $V <$ 10 from the
Yerkes 1.0m \citep{Schlesinger(1911)}.  Today, the two cornerstones of
parallax results are {\it The Catalogue of Trigonometric Parallaxes,
  Fourth Edition} \citep{vanAltena(1995)} from Yale University
Observatory, often called the {\it Yale Parallax Catalog} (hereafter
referred to as YPC) and the {\it HIPPARCOS} catalog
(\citealt{Perryman(1997)}, updated in \citealt{vanLeeuwen(2007)};
hereafter referred to as HIP), with many of the brighter YPC M dwarfs
also having been observed by HIP.  Most of the parallaxes for nearby M
dwarfs in the YPC are results from the long-term astrometric programs
that were carried out with telescopes at the Allegheny, Cape,
McCormick, Mt. Wilson, Sproul, Van Vleck, Yale, Yerkes and US Naval
Observatories. The $\sim$475 nearby M dwarf systems included in HIP
are bright ($V$ $\lesssim$ 12) and mostly early-type objects, while
the $\sim$650 nearby M dwarfs listed in the YPC encompass a wide range
of M dwarf spectral types. The faintest M dwarf included in YPC
  (\object{RGO 0050-2722}, with $V =$ 21.51 \citep{Ianna(1993)}) is
  due to astrometrists' continuing efforts to push the faintness
  limits of parallax measurements \citep{Ianna(1986)}.


Recent interest in M dwarfs as targets for exoplanet searches has
spurred parallax measurements for these stars at the bottom of the
stellar main-sequence.  Many astrometric programs
\citep{Gatewood(2008), Gatewood(2009), Lepine(2009), Smart(2007),
  Smart(2010), Dupuy(2012), Shkolnik(2012), Dittmann(2014),
  Finch(2016)} have contributed important distance measurements to the
nearby M dwarf population, primarily in the northern hemisphere.
Additional results have become available from re-reductions of the
{\it Hipparcos} data, e.g., \citet{Soderhjelm(1999)} and
\citet{Fabricius(2000)}.


The REsearch Consortium On Nearby Stars' (RECONS)\footnote{\it
  www.recons.org} work over the past two decades has helped fill in
the nearby star census with vital distance determinations,
specifically in the historically neglected southern hemisphere.  To
date, RECONS has contributed the first reliable $\pi_{trig}$
measurements for $\sim$200 M dwarfs, as well as for white dwarfs and
brown dwarfs, within 25 pc via our long-term astrometric program at
the CTIO/SMARTS 0.9m and 1.5m
\citep{Costa(2005),Costa(2006),Jao(2005),Jao(2011),Jao(2014),Gizis(2007),
  Henry(1997),Henry(2006),Subasavage(2009),Riedel(2010),Riedel(2011),
  Riedel(2014),vonBraun(2011),Mamajek(2013),Dieterich(2014),Lurie(2014),
  Davison(2015)}.  The work presented here is part of the continuing
RECONS effort to determine a complete census of all M dwarfs within 25
pc.  Here we present the largest set of accurate $\pi_{trig}$
measurements for stars in the southern hemisphere since the YPC and
HIP.


\section{Definition of the Sample}
\label{sec:sampledef}

Stars reported here were targeted during the astrometry program
because they were likely to be red dwarfs within 25 pc with no
previous published $\pi_{trig}$.  Of the 151 systems, 93 are from
previous compendia of proper motion stars, primarily based on work by
Luyten
\citep{Luyten(1979a),Luyten(1979b),Luyten(1980a),Luyten(1980b)}, and
58 are from our SuperCOSMOS-RECONS (SCR) search
\citep{Hambly(2004),Henry(2004)}.  The systems have $\mu$ = 118--828
mas yr$^{-1}$, with 143 having $\mu$ $>$ 180 mas yr$^{-1}$, the
canonical cutoff for Luyten's Two-Tenths (LTT) Catalog.

We define M dwarfs to have 3.7 $\leq$ $(V-K)$ $\leq$ 9.5 and 8.8
$\leq$ $M_{V}$ $\leq$ 20.0, corresponding to objects with masses
0.64--0.07 M$_{\odot}$ \citep{Benedict(2016)}. For the most massive M
dwarfs, these color and absolute magnitude limits were determined at
the K/M dwarf boundary by creating observational HR diagrams from the
RECONS 25 Parsec Database (see \S \ref{sec:future}), where spectral
types from RECONS' work, \citet{Gray(2003)}, \citet{Reid(1995)}, and
\citet{Hawley(1996)} were used to separate the K and M dwarfs.
Because spectral types can be imprecise, there was overlap between the
two spectral types; boundaries were thus chosen to split the types at
carefully defined $(V-K)$ and $M_V$ values.  A similar method was
followed for the M/L dwarf transition using results from both RECONS
and \citet{Dahn(2002)}, providing the regimes outlined above for the
observational luminosities and colors that include the full range of M
dwarfs. Stars for which astrometry and photometry are presented here
have 3.85 $<$ $(V-K)$ $<$ 8.47 and are therefore all M dwarfs.



Figure \ref{fig:sample_def} provides some of the general
characteristics of the stars presented here. The left panel shows the
distribution on the sky for the sample (note that there are eight
stars with $\delta$ $>$ 0$^{\circ}$).  The right panel illustrates an
observational luminosity function, using $M_V$, for the sample,
indicating that the majority of these M dwarfs are of the mid-M
spectral type. The true distribution would shift slightly toward
  lower mass stars with the deblending of the photometry of the
  twenty-one unresolved binaries (indicated in blue).




\section{Observations \& Data Reduction}
\label{sec:data}

\subsection{Photometry}
\label{subsec:ccd.phot}

\subsubsection{$V_JR_{KC}I_{KC}$ CCD Photometry} 
\label{subsubsec:vriphot}

For all but two objects,\footnote{Only one epoch of photometry was
  available for WT 1637 and LHS 2024 at the time of this paper.} at
least two epochs of absolute $V_JR_{KC}I_{KC}$ (henceforth referred to
as $VRI$) photometry on the Johnson-Kron-Cousins system were measured
for each parallax field.  Two $V$ filters that are photometrically
indistinguishable to 7 millimagnitudes (mmag) \citep{Jao(2011)}, one
$R$ filter, and one $I$ filter were used for series of observations
spanning 2--16 years, depending on the star.\footnote{The central
  wavelengths for the two $V_J$ filters, the R$_{KC}$ filter, and the
  I$_{KC}$ filter are 5438\AA, 5475\AA, 6425\AA, and 8075\AA,
  respectively.}  The 2048 $\times$ 2046 Tektronix CCD camera on the
CTIO/SMARTS 0.9m with a pixel (px) scale of 0\farcs401 px$^{-1}$ was
used for both astrometric and photometric observations.  In order to
mitigate the effects of image distortion at the edges of the CCD, only
the central quarter of the chip was used, resulting in a 6\farcm8
$\times$ 6\farcm8 square field of view.

The data were reduced using $IRAF$ \footnote{$IRAF$ is distributed by
  the National Optical Astronomy Observatories, which are operated by
  the Association of Universities for Research in Astronomy, Inc.,
  under cooperative agreement with the National Science Foundation.}
\citep{Tody(1993)}.  Bias subtraction and flat fielding were done with
calibration frames acquired nightly.  Most of the targets were
observed at {\it sec z} $<$ 1.8, although for a few of the more
extreme northern stars, this was not possible.  Standard stars from
\citet{Bessel(1990)},
\citet{Landolt(1992),Landolt(2007),Landolt(2013)} and/or
\citet{Graham(1982)} were observed multiple times throughout the
photometry nights at varying airmasses and used to determine
transformation equations and extinction curves.  Apertures
14$\arcsec$~in diameter were used to determine the stellar fluxes,
matching the aperture diameters used by Landolt.  In cases where close
contaminating sources required deblending, smaller apertures were used
and aperture corrections were applied.

A few systems had separations too small to allow aperture corrections;
thus, point spread function (PSF) photometry was performed using
either $IRAF$ or $IDL$. For SCR 1630-3633AB, the standard $IRAF$ PSF
reduction procedures\footnote{as outlined in {\it A User's Guide to
    Stellar CCD Photometry with IRAF} from NOAO} were followed to
obtain individual photometry for the components of these close
binaries. For all other PSF reductions, a custom $IDL$ program was
used. In short, the contribution from the sky background was
calculated from a `blank' part of the image. The region around the
close pair being analyzed was then cropped to include only the
relevant pair, and the background subtracted. A Moffat curve was
  fit to the PSF of the primary, the flux determined from the fit, and
  then the primary was subtracted from the image. Care was taken to
  minimize the residual counts from the primary. Gaussian and
  Lorentzian curves were also tested, but it was found that the Moffat
  curve fit the shape of the PSF best. A Moffat curve was then fit to
  the secondary component's PSF and the flux calculated from the fit.
The ratio of the fluxes then yielded the $\Delta$mags from which
individual magnitudes were calculated from the blended
photometry. $VRI$ photometry from PSF reductions listed in Table
\ref{tab:photometry} are noted.

As outlined in \citet{Winters(2011)}, photometric uncertainties are
typically 30 mmag for the $V-$band and 20 mmag for the $R-$ and
$I-$bands.  Further details on the photometric data reduction, the
transformation equations, uncertainties, etc., can be found in
\citet{Jao(2005),Jao(2011),Winters(2011)} and \citet{Winters(2015)}.
Note that a `u' next to the reference in column (7) indicates that
photometry was reported in a previous RECONS publication, but that the
value listed here is updated due to the acquisition of additional
data.


\subsubsection{$JHK_s$ Photometry from 2MASS}
\label{subsubsec:IRphot}

Photometry in the near-infrared $JHK_s$ (hereafter simply $JHK$)
filters has been extracted from 2MASS \citep{Skrutskie(2006)} and is
rounded to the nearest hundredth magnitude in Table
\ref{tab:photometry}.  Uncertainties in $JHK$ are typically less than
0.05 mag, with exceptions indicated in columns (8,9,10) of Table
\ref{tab:photometry}.

All $JHK$ magnitudes have been confirmed by eye to correspond to the
correct objects by using the Aladin interface at {\it CDS} to `blink'
SuperCOSMOS images taken at different epochs and identifying the 2MASS
point that matches the projected proper motion.  Because the proper
motions of all the objects presented here are larger than 118 mas
yr$^{-1}$, their movements were easily discernible during the blinking
process and the correct identification of the 2MASS source was
straightforward.


Table \ref{tab:photometry} provides the $VRIJHK$ photometry data for
the 160 stars in 151 M dwarf systems presented here.  Included are the
names of the M dwarfs (column 1), coordinates (J2000.0) (2,3), $VRI$
magnitudes (4,5,6) and the number of observations and/or reference
(7), the 2MASS $JHK$ magnitudes (8,9,10), and a photometric distance
estimate and uncertainty based on the photometry (11,12).  Components
of multiple systems are noted with capital letters A and B after the
names in the first column.  If the names of the components are
different, the letters identifying the primary and the secondary are
placed within parentheses (e.g., LTT 7419(A) and CE 507(B), where the
`B' component is also known as LTT 7419B).  A 'J' next to a magnitude
indicates that light from a close companion has resulted in blended
photometry.  A note has been added to the photometric distance
estimates for the 27 systems for which the trigonometric distance
places the system beyond 25 pc.  Nearly all of these stars were
originally thought to be within 25 pc, but our new astrometry
indicates that they are not.  This in itself reveals important
information about the stars, as discussed in $\S$4.1.



\begin{deluxetable}{lccccccccccc}
\centering
\setlength{\tabcolsep}{0.03in}
\tablewidth{0pt}
\tabletypesize{\scriptsize}
\tablecaption{Photometric Data}
\tablehead{\colhead{Name}                &
	   \colhead{RA}                  &
 	   \colhead{DEC}                 &
	   \colhead{$V_J$}               &
           \colhead{$R_{KC}$}            &
	   \colhead{$I_{KC}$}            &
	   \colhead{\# obs/ref}          &
           \colhead{$J$}                 &
	   \colhead{$H$}                 &
           \colhead{$K_s$}               &
           \colhead{d$_{CCD}$}           &
           \colhead{$\sigma_{CCD}$}      \\

	   \colhead{   }                 &
	   \colhead{(hh:mm:ss)}          &
           \colhead{(dd:mm:ss)}          &
           \colhead{(mag)}               &
           \colhead{(mag)}               &
	   \colhead{(mag)}               &
           \colhead{     }               &
           \colhead{(mag)}               &
           \colhead{(mag)}               &
           \colhead{(mag)}               &
           \colhead{(pc)}                &
           \colhead{(pc)}                \\

	   \colhead{(1)}                 &
	   \colhead{(2)}                 &
           \colhead{(3)}                 &
           \colhead{(4)}                 &
           \colhead{(5)}                 &
	   \colhead{(6)}                 &
           \colhead{(7)}                 &
           \colhead{(8)}                 &
           \colhead{(9)}                 &
           \colhead{(10)}                &
           \colhead{(11)}                &
           \colhead{(12)}                }

\startdata
\object[SCR J1630-3633AB]{SCR 1630-3633A}        &16 30 27.2  &$-$36 33 56  & 15.35\tablenotemark{a}   & 13.83   & 11.92   &  1/1   &  10.04J                      &   9.50J                    &   9.03J                   & \nodata               &\nodata \\ 
SCR 1630-3633B                                   &16 30 27.2  &$-$36 33 56  & 19.98\tablenotemark{a}   & 19.17   & 18.26   &  1/1   &   \nodata                    &   \nodata                  &   \nodata                 & \nodata               &\nodata \\ 
SCR 0644-4223A                                   &06 44 32.1  &$-$42 23 45  & 14.74\tablenotemark{a}   & 13.47   & 11.95   &  1/2u  &   9.93J                      &   9.27J                    &   8.98J                   & \nodata               &\nodata \\ 
SCR 0644-4223B                                   &06 44 32.1  &$-$42 23 45  & 16.04\tablenotemark{a}   & 14.77   & 12.82   &  1/2u  &   \nodata                    &   \nodata                  &   \nodata                 & \nodata               &\nodata  \\ 
L 345-91B                                        &18 48 41.4  &$-$46 47 08  & 12.45\tablenotemark{a}   & 11.33   & 10.06   &  1/3u  &   \nodata                    &   \nodata                  &   \nodata                 & \nodata               &\nodata  \\  
LP 382-56A                                       &06 57 11.7  &$-$43 24 51  & 11.89\tablenotemark{a}   & 10.85   &  9.47   &  1/3u  &   7.59J                      &   6.98J                    &   6.76J                   & \nodata               &\nodata  \\
LP 382-56B                                       &06 57 11.7  &$-$43 24 51  & 11.95\tablenotemark{a}   & 10.89   &  9.81   &  1/3u  &   \nodata                    &   \nodata                  &   \nodata                 & \nodata               &\nodata \\
\enddata

\tablenotetext{a}{$VRI$ photometry for individual components has been
  calculated using PSF reduction.}

\tablenotetext{b}{$2MASS$ magnitude uncertainty greater than 0.05
  mags.}

\tablenotetext{c}{The $\pi_{trig}$ for this object places it beyond 25
  pc.}

\tablenotetext{d}{The photometric distance estimate is based upon the
  joint photometry for the system and is thus likely an
  underestimate.}

\tablecomments{A `J' next to a photometry value indicates that the
  magnitude is blended due to one or more close companions.  A `u'
  next to the reference indicates that RECONS photometry was reported
  in a previous publication, but that the value listed here is updated
  due to additional data and supersedes previous results.}

\tablerefs{(1) this work; (2) \citet{Winters(2011)}; (3)
  \citet{Winters(2015)}.\label{tab:photometry}}

\end{deluxetable}



\clearpage

\subsection{Astrometry}
\label{sec:astr}

Here we briefly outline the methods used for astrometry observations
and data reduction.  Parallax frames are taken using either $V_J$
(old), $V_J$ (new)\footnote{Between 2005 March and 2009 September, the
  $V_J$ (new) filter was used in lieu of the $V_J$ (old) filter and
  was found to be slightly inferior, astrometrically; however, as
  noted in \citet{Subasavage(2009)}, there is a benefit to including
  data taken with both filters if there are $\sim$ 1-2 years of data
  taken after the filter switch. See \citet{Subasavage(2009)} and
  \citet{Jao(2011)} for a more thorough discussion. Of the targets
  presented here, 62 were observed in the $V$ filter; all but two ---
  L 749-34 and LHS 3124 --- have either all of their observations in
  the $V_J$ (old) filter or the necessary 1-2 years of additional data
  after the filter switch. Neither L 749-34 nor LHS 3124 shows any
  signs of perturbation, so it is not felt that the astrometry has
  been affected.}, $R_{KC}$, or $I_{KC}$ filters, depending on the
brightness of the science target, using the same instrumental set-up
as that used for the photometry frames.  The magnitudes of the stars
span 7.94 $\leq$ $VRI$ $\leq$ 18.55.  To balance the signal from the
red target stars to generally fainter, bluer, reference stars, bright
stars were observed in $V$, while to increase the signal for faint
target stars, the $I$ filter was used.  This reduces the magnitude
range for target star observations in the parallax filters to 9.74
$\leq$ $VRI$ $\leq$ 15.91.

Typically, three to five frames (depending on the target's brightness)
were taken each night within 30 minutes of the science target's
meridian transit to minimize corrections required for differential
color refraction (DCR).  The parallax star and the reference stars do
not have identical colors; therefore, their positions shift relative
to each other as a result of refraction by the Earth's atmosphere,
necessitating the DCR corrections.  Mean exposure times ranged from 20
to 1200 seconds, depending on the brightnesses of the science stars,
the reference field stars, and sky conditions.  All data were taken
between 1999 and 2016, with timespans for observations for each star
given in column (8) of Table \ref{tab:ctiopi_data}.  As with the
photometry frames, bias and dome flat frames were taken nightly to
enable calibration of the science images, following the procedures
outlined in \citet{Jao(2005)}.

As discussed extensively in \citet{Jao(2005)}, \citet{Henry(2006)},
and \citet{Subasavage(2009)}, our data reduction processes disentangle
the proper motions and parallactic shifts that contribute to the paths
traced on the sky by the science targets.  In short, (1) {\it
  SExtractor} \citep{Bertin(1996)} was used to measure the centroids
of the science and reference stars, (2) a six-constant plate model was
defined to determine plate constants in order to rotate and translate
each frame so that the reference stars were aligned from frame to
frame, (3) least-square equations were then solved for multi-epoch
series of images using {\it Gaussfit} \citep{Jefferys(1988)}, and (4)
the relative parallax was corrected to absolute parallax via the
photometric distance estimates of the reference stars, which were
initially assumed to have zero mean parallax and proper motion. 

As noted in \citet{Subasavage(2009)}, in 2007, our reduction processes
began incorporating a new centroiding algorithm available in the {\it
  SExtractor} package (see the {\it SExtractor} v2.13 User's
manual\footnote{https://www.astromatic.net/pubsvn/software/sextractor/trunk/doc/sextractor.pdf}
for a more thorough discussion) that utilizes a circular
Gaussian-weighted two-dimensional window to iteratively measure the
pixel coordinates of the reference and parallax stars. The $x$ and $y$
pixel coordinates are measured through the $XWIN_{IMAGE}$ and
$YWIN_{IMAGE}$ outputs. The resulting positional parameters are much
more precise than those obtained using the isophotal method and are
very close to the precision obtained using a PSF-fitting model.

The final $\pi_{trig}$ uncertainty is a combination of many factors,
including (1) the quality of the reference star frame (number of
stars, distribution, brightness), (2) the accuracy of the $(x, y)$
centroids, including any ellipticity caused by a close component, (3)
the total number of images used in the $\pi_{trig}$ measurement, (4)
the time span of the images used, (5) the parallax factor coverage
(i.e., the breadth of parallax factors covered and how many frames
were obtained for the `morning' and `evening' halves of the parallax
ellipse for each field), (6) the DCR corrections, and (7) the
correction from relative to absolute parallax.

Table \ref{tab:ctiopi_data} presents details of the astrometry
observations, including the key trigonometric parallax results
($\pi$), as well as proper motions ($\mu$) and derived tangential
velocities ($V_{tan}$).  After the names (column 1), we list J2000.0
coordinates (2,3), the filter in which $\pi_{trig}$ was measured (4),
the number of seasons N$_{sea}$ covered\footnote{N$_{sea}$ indicates
  the number of seasons observed, where the six months when a star is
  observable counts as one season. The letter `c' indicates a
  continuous set of observations where multiple nights of data were
  taken in each season, whereas an `s' indicates scattered
  observations when some seasons have only a single night of
  observations.} (5), the number of frames N$_{fr}$ acquired for each
star (6), the time interval covered by the data sequence (7), the
number of years of coverage (8), and the number of reference stars
N$_{ref}$ used in the field (9).  Results for photometric variability
in the filter used for parallax observations are given in column (10),
followed by the relative parallax $\pi$$_{rel}$ (11), the correction
to absolute parallax $\pi$$_{cor}$ (needed because the reference stars
are not at an infinite distance) (12), and final absolute parallax
$\pi$$_{abs}$ (13), all with their associated uncertainties.  The
magnitude and position angle (measured east from north) of the proper
motion, with their uncertainties, are given in columns (14,15),
followed by the derived tangential velocity $V_{tan}$ (16), and notes
(17).  Objects with `!'  in the notes column are discussed further in
Section \ref{subsec:indiv}.

\section{Results}
\label{sec:results}

\subsection{Photometric Distance Estimates}
\label{subsec:ccddist}

Because $\pi_{trig}$ measurements require significant resources in
terms of both time and money, a careful sequence of events takes place
before an object is targeted for astrometry.  Generally, before
placing a potentially nearby star on the program, $VRI$ magnitudes are
measured and $JHK$ magnitudes are extracted from 2MASS.  Photometric
distance estimates are then calculated using the relations described
in \citet{Henry(2004)}.  Of the 15 possible color--$M_K$ relations, 12
are predictive and are calculated using the available
photometry.\footnote{Only two of the stars offered here have fewer
  than 12 relations: GJ 714 has 10 relations and LP 349-25AB has 11
  relations.}  These are then compared to the same relations for known
single stars with accurate $\pi_{trig}$.  Any objects with distance
estimates placing them closer than 20 pc are added to the program. The
distance estimates and uncertainties are provided in the final two
columns of Table \ref{tab:photometry}.  The few stars presented here
with distance estimates placing the beyond 20 pc are a mixture of
those with distances estimated by other investigators to be nearby,
relatively high proper motion stars, and stars added soon after the
program started to fill in hours of RA that had few targets.

For six close binary systems, the determination of individual
parallaxes for each component was possible, as was the deblending of
their $VRI$ photometry via PSF reductions; however, the 2MASS
photometry is unresolved. These systems have been omitted from Figures
\ref{fig:ccd.trig}, \ref{fig:hr}, and \ref{fig:var.vk}, as there is no
clean way to include them.



Most of the stars in Table \ref{tab:photometry} were published
previously in \citet{Winters(2011)} or \citet{Winters(2015)}.
Exceptions include 16 stars.  The sample presented in
\citet{Winters(2015)} were southern M dwarf primaries; thus, neither
the eight northern M dwarfs nor the seven secondary components
presented here were included in that publication.  One additional
southern M dwarf primary (SCR 2303-4650) was overlooked in previous
publications and is offered here for the first time.

We note that the distance estimates reported here may be slightly
different from those reported previously, due to the acquisition of
more $VRI$ photometry that improves the estimate.  These new results
thus supersede previous results.

\subsection{Trigonometric Distance Measurements}
\label{subsec:trigdist}


The astrometry results presented in Table \ref{tab:ctiopi_data}
represent the largest set of $\pi_{trig}$ measurements with
uncertainties less than $\sim$3 mas (milliarcseconds) measured for
stars in the southern hemisphere since YPC and HIP.  The distribution
of the $\pi_{trig}$ uncertainties in Figure \ref{fig:pi_err}
illustrates that most of the measurements have errors less than 2 mas,
with a mean value of 1.6 mas.  A large $\pi_{trig}$ error is usually
caused by a sparse or asymmetric reference field or the presence of a
close (gravitationally bound or background) star.  We note that
targets in the direction of the Galactic center will have reddened
reference stars that corrupt the correction from relative to
absolute parallax because their distance estimates are too close.  For
these objects, we have adopted generic corrections of 1.50 $\pm$ 0.50
mas, as noted in Table \ref{tab:ctiopi_data}.


Comparisons of the photometric distance estimates to the accurate
trigonometric distances provide valuable information about the
targets. Unresolved binaries will have underestimated photometric
distances, due to the added light from unseen companion(s). Unresolved
equal-magnitude binaries will have distance estimates underestimated
by a factor of sqrt(2).  Light from fainter companions will decrease
this offset, whereas light from additional unresolved companions will
increase this offset.  Low-mass subdwarfs will have overestimated
distances, as they are underluminous at a given color; however, no
known subdwarfs are included in the sample presented here.

Figure \ref{fig:ccd.trig} compares the photometric distance estimates
to the measured trigonometric distances, with the solid line tracing
distances that match.  A group of points with larger trigonometric
distances than photometric estimates is evident, primarily due to a
number of both confirmed and candidate unresolved binaries.  The
dash-dot line indicates the boundary at which the trigonometric
distance is larger than sqrt(2) times that of the photometric
distance, corresponding to an equal-luminosity pair of unresolved
stars.  Points for candidate binaries below this line, taking into
account the uncertainties on their photometric distances, are enclosed
with open red squares, whereas points for known unresolved multiples
are enclosed with red stars.  Alternately, some of the stars below the
dash-dot line may be young stars that are overluminous because they
have inflated radii, resulting in underestimated photometric distance
estimates.  However, we expect the number of unidentified young stars
in this sample to be small because we have vetted the entire observing
list for young stars with various attributes of youth, e.g., x-ray
emission and space motion that matches known young moving groups.  The
inset histogram shows the distribution of the absolute deviations
between the two distances.  For the presumed single stars within 25
pc, the mean absolute deviation is 21\%, a bit higher than the 15\%
reported in \citet{Henry(2004)}, the offset outlined with dashed
lines.  This is due to the presence of the candidate binaries
mentioned above.  Both the known and candidate multiples are discussed
in $\S$5.






\begin{deluxetable}{lcccccccccccccccc}
\rotate
\setlength{\tabcolsep}{0.03in}
\tablewidth{0pt}
\tabletypesize{\tiny}
\tablecaption{Astrometry Data}
\tablehead{\colhead{Name}                &
	   \colhead{RA}                  &
 	   \colhead{DEC}                 &
	   \colhead{fil}                 &
	   \colhead{N$_{sea}$}           &
           \colhead{N$_{fr}$}            &
           \colhead{coverage}            &
           \colhead{T}                   &
           \colhead{N$_{ref}$ }          &
           \colhead{var}                 &
           \colhead{$\pi$$_{rel}$}       &
           \colhead{$\pi$$_{cor}$}       &
           \colhead{$\pi$$_{abs}$}       &
           \colhead{$\mu$}               & 
           \colhead{P.A.}                &
           \colhead{V$_{tan}$}           &
           \colhead{notes}               \\

	   \colhead{   }                 &
	   \colhead{(hh:mm:ss)}          &
           \colhead{(dd:mm:ss)}          &
           \colhead{   }                 &
           \colhead{   }                 &
           \colhead{   }                 &
           \colhead{   }                 &
           \colhead{(yr)}                &
           \colhead{   }                 &
           \colhead{(mmag)}              &
           \colhead{(mas)}               &
           \colhead{(mas)}               &
           \colhead{(mas)}               &
           \colhead{(mas yr$^{-1}$)}      &
	   \colhead{(deg)}               &
           \colhead{(km s$^{-1}$)}        &
           \colhead{}                    \\

	   \colhead{(1)}                 &
	   \colhead{(2)}                 &
           \colhead{(3)}                 &
           \colhead{(4)}                 &
           \colhead{(5)}                 &
	   \colhead{(6)}                 &
           \colhead{(7)}                 &
           \colhead{(8)}                 &
           \colhead{(9)}                 &
           \colhead{(10)}                &
           \colhead{(11)}                &
           \colhead{(12)}                &
           \colhead{(13)}                &
	   \colhead{(14)}                &
	   \colhead{(15)}                &
           \colhead{(16)}                &
           \colhead{(17)}                }

\startdata
L 217-28        & 00 08 17.37 & $-$57 05 52.9  & V &   8s &  68  & 2000.50-2008.00 &   7.25 &   7  &   9.8 &  73.66 $\pm$ 2.10 & 1.51 $\pm$ 0.21                  &  75.17 $\pm$ 2.11                   &  385.4 $\pm$ 0.7 &  263.1 $\pm$ 0.17 &  24.3 &    \\ 
LEHPM 1-255     & 00 09 45.06 & $-$42 01 39.6  & V &   4c &  51  & 2009.75-2012.94 &   3.20 &   6  &  13.7 &  52.12 $\pm$ 1.49 & 1.14 $\pm$ 0.22                  &  53.26 $\pm$ 1.51                   &  254.8 $\pm$ 1.4 &  091.4 $\pm$ 0.44 &  22.7 &    \\
SCR 0027-0806   & 00 27 45.36 & $-$08 06 04.7  & I &   5s &  59  & 2008.85-2014.94 &   6.09 &   9  &  13.3 &  23.39 $\pm$ 1.24 & 1.50 $\pm$ 0.10                  &  24.89 $\pm$ 1.24                   &  118.5 $\pm$ 0.9 &  118.6 $\pm$ 0.80 &  22.6 &    \\
LP 349-25AB     & 00 27 55.99 & $+$22 19 32.3  & I &  12s & 100  & 2003.51-2015.92 &  12.42 &   7  &  11.2 &  67.71 $\pm$ 1.39 & 1.45 $\pm$ 0.14                  &  69.16 $\pm$ 1.40\tablenotemark{a}  &  429.8 $\pm$ 0.3 &  112.4 $\pm$ 0.07 &  29.5 & !  \\
LTT 313         & 00 35 38.07 & $-$10 04 18.7  & V &   6s &  70  & 2010.97-2015.97 &   5.00 &   7  &   9.9 &  31.14 $\pm$ 1.32 & 1.38 $\pm$ 0.08                  &  34.52 $\pm$ 1.32\tablenotemark{a}  &  211.1 $\pm$ 1.0 &  254.4 $\pm$ 0.46 &  30.1 & !  \\
LHS 1134        & 00 43 26.01 & $-$41 17 34.0  & V &   7s &  72  & 2009.78-2015.97 &   6.19 &   8  &  10.5 &  94.68 $\pm$ 1.40 & 0.56 $\pm$ 0.08                  &  95.24 $\pm$ 1.40                   &  764.4 $\pm$ 0.8 &  220.1 $\pm$ 0.11 &  38.0 &    \\
LHS 1140        & 00 44 59.34 & $-$15 16 17.5  & V &   8s &  60  & 2003.95-2014.93 &  10.98 &   5  &  12.5 &  79.57 $\pm$ 2.76 & 0.61 $\pm$ 0.11                  &  80.18 $\pm$ 2.76                   &  665.9 $\pm$ 1.0 &  153.3 $\pm$ 0.16 &  39.4 &    \\
\enddata

\tablenotetext{a}{Because this field is in the direction of the
  Galactic center, the initial correction from the relative parallax
  to the absolute parallax for this object was erroneously large due
  to reddened reference stars.  We have adopted a generic correction
  of 1.50 $\pm$ 0.50 mas.}

\tablenotetext{b}{These parallax values have had the orbital motion of
  the system subtracted. \label{tab:ctiopi_data}}

\end{deluxetable}
\clearpage

\subsection{Stars on the HR Diagram}
\label{subsec:hrd}

Figure \ref{fig:hr} is an observational HR diagram illustrating the
stars presented here, using $M_V$ for luminosity and $V-K$ for
temperature.  The small black points are single stars from the RECONS
25 Parsec Database (see $\S$\ref{sec:future}) for reference, while the
stars with new trigonometric distances are shown as larger red points.
Multiples, both known (stars), and candidates (squares) are indicated,
and described in $\S$5.  The candidates were identified based on their
positions above the main sequence on this plot and from mismatches
between their photometric distance estimates and their trigonometric
distances shown in Figure \ref{fig:ccd.trig}.  As described in
$\S$5.4, these objects are overluminous for their distances and we
thus expect them to be either near-equal luminosity and mass binaries,
or young.  The new multiples described in $\S$5.1 and $\S$5.2 whose
positions on this diagram are within the main sequence are expected to
have unseen companions that contribute little light to the systems.
Two of the new systems whose positions are slightly below the main
sequence (SCR 0723-8015AB and LTT 9828AB) may have white dwarf
companions, as the magnitudes of their perturbations are quite large,
and yet they are not elevated above the main sequence, nor is there a
mismatch between their photometric and trigonometric distances.

\subsection{Stellar Variability}
\label{subsec:var}

We also investigate the photometric variability of stars observed
during the program using the astrometry frames to reveal short- or
long-term photometric changes, such as flares or stellar cycles.
Results are given in column (10) of Table \ref{tab:ctiopi_data}.  In
short, the instrumental magnitudes of the parallax stars are compared
to the ensemble of reference stars, with corrections made for varying
integration times, seeing differences, etc., between frames in a
series.  Reference stars exhibiting variability are removed from the
analysis.  Details on the methodology can be found in
\citet{Jao(2011)} and \citet{Hosey(2015)}.


We find that most of the stars presented here vary by less than the 20
mmag (2\%) level we defined in \citet{Hosey(2015)} to be the threshold
for significantly variable stars.  This threshold is based on our
measurement limit of $\sim$7 mmag, which we adopt as the 1$-$$\sigma$
minimum deviation threshold for the carefully focused images required
for the astrometry program.\footnote{Four stars have variability
  measured to be slightly less than 7 mmag, at 6.4--6.9 mmag.}  The
limit can, and has been, reduced to 2--3 mmag on the 0.9m using
defocused images and longer integrations.  The 15 stars without
variability measurements in Table \ref{tab:ctiopi_data} (a) are close
multiples separated by a few arcseconds that are photometrically
entangled, (b) were corrupted by a background star during many images
in the series, or (c) were underexposed relative to the primary star
so that the variability measurement was deemed unreliable.


Six stars were variable by more than 20 mmag, with L 2-77 exhibiting
the largest value at 34.0 mmag.  The other variable stars are LTT
7434AB (28.5 mmag), SCR 1214-4603 (27.3 mmag), L 532-12 (25.2 mmag),
LEHPM 1-1343 (24.2 mmag), and L 204-148 (20.9 mmag).  Figure
\ref{fig:var.vk} illustrates variability of the targets in their
respective parallax filters ($VRI$) as a function of their $M_V$.
Unexpectedly, we do not see any noticeable increase in variability in
the bluest, $V$, filter.  We note that three of the six variable
stars (above the 20 mmag line) were observed in the $R-$band, but
based on the small number of variable stars, we do not feel that any
trend is evident.  We do not see any relation between the presence of
an unresolved stellar companion and the magnitude of the variability,
although our astrometric series are not sensitive to very close
systems with orbital periods of a few days in which tidal effects may
cause stellar activity.  Overall, the results for the 145 stars
presented here agree with the results presented in
\citet{Hosey(2015)}, with an overlap of 44 stars between the two
samples.

We highlight four of the most interesting variable objects
here. Figure \ref{fig:var_weirdos2} highlights two of the most
variable stars in the sample. In the left panel, SCR 1214-4603 shows
evidence of a flare event in 2011 with a temporal coverage of eight
years. In the right panel is shown L 2-77 with almost four years of
coverage. This star was fairly stable through 2012, but has recently
started showing signs of flares and spots.

Shown in the left panel of Figure \ref{fig:var_weirdos} is the clear
stellar cycle of SCR 0702-6102, with a period on the order of a
decade.  The right panel shows the long-term increasing brightness of
G 169-29, originally presented in \citet{Hosey(2015)}.  Five
additional years of data are available for analysis, confirming the
brightening trend that now extends to 15 years.  Note that the
variability levels for these stars --- 9.8 mmag for SCR 0702-6102 and
17.8 mmag for G 169-29 --- do not meet our threshold for significant
variability, although it is clear that both show long-term changes in
brightness.  As we accumulate more data for hundreds of stars, we will
explore additional facets of the long-term variability of red dwarfs
that occur below 20 mmag.


\subsection{Notes on Individual Systems}
\label{subsec:indiv}

Here we present details for some of the more interesting or
complicated systems, ordered by RA.  The first four digits of both the
RA and DEC are provided in parentheses after the names, and Figures
are listed if plots of the astrometry series are included.  

\hspace{-22pt}{\bf LP 349-25AB (0027+2219), Figure
  \ref{fig:perturbations1}:} This star shows a perturbation in both
axes.  With over ten years of data, the orbit appears to have recently
wrapped in our astrometry observations. The system is overluminous,
resulting in a photometric distance estimate of 8.2 pc, compared to
its trigonometric distance of 15.4 pc, implying that the companion
contributes significant flux to the system in the $I$-band. We note
that \citet{Forveille(2005)} report a companion detected via AO in
2004 with $\rho$ $=$ 0.107\arcsec~and $\theta$ $=$ 7$^{\circ}$, with
$\Delta$$H$ $=$ 0.4 mag.


{\hspace{-22pt}{\bf LTT 313 (0035$-$1004):} The trigonometric distance
  for this star is about twice its photometric distance (30.0 $\pm$
  1.2 pc versus 17.5 $\pm$ 2.7 pc).  We detect a hint of a
  perturbation in the DEC axis (not shown), but at a distance $>$ 25
  pc, we are no longer following this star to confirm or refute the
  possible perturbation.


\hspace{-22pt}{\bf L 294-92AB (0147-4836), Figure
  \ref{fig:perturbations1}:} This star shows a perturbation in both
axes, although both types of distances match within the uncertainties:
11.5$\pm$ 1.8 pc for the photometric distance versus 13.8 $\pm$ 0.3 pc
for the trigonometric distance.  Thus, the companion is likely of low
mass and contributes little light to the system in the $R-$band.  The
orbit has not yet wrapped in our astrometric series.



\hspace{-22pt}{\bf L 88-43 (0153$-$6653):} The large uncertainty on
the $\pi_{trig}$ for this object is due to its weak reference field
and short integrations, typically only 15--30 seconds in $R$.
 
\hspace{-22pt}{\bf LP 831-45AB (0314-2309), Figure
  \ref{fig:perturbations1}:} This system exhibits a perturbation, very
obvious in DEC, resulting in a $\sim$6 year period for the pair.  The
distances are 12.8 $\pm$ 2.0 pc for the photometric distance versus
15.7 $\pm$ 0.3 pc for the trigonometric distance, implying that this
is an unequal mass binary.

\hspace{-22pt}{\bf L 591-42AB (0436-2721):} While one component lies
within 25 pc (A has $\pi$ $=$ 42.02 $\pm$ 2.48 mas), and the other
beyond (B has $\pi$ $=$ 38.98 $\pm$ 2.59 mas), these results agree
within the somewhat large uncertainties.  The weighted mean
$\pi_{trig}$ is 40.57 $\pm$ 1.79 mas, placing it just within 25 pc.
In 2001, the two stars were separated by $\sim$50\arcsec~at a position
angle of $\theta$ $=$ 196$^{\circ}$ \citep{Jao(2003)}.


\hspace{-22pt}{\bf SCR 0702-6102 (0702-6102), Figure
  \ref{fig:var_weirdos}:} The discrepant distances for this star (10.7
$\pm$ 2.0 pc for its photometric distance versus 16.8 $\pm$ 0.4 pc for
its trigonometric distance) and its elevation above the main sequence
in the HR diagram in Figure \ref{fig:hr} suggest that this is an
unresolved binary or a young star.  The astrometric data show a hint
of a perturbation with an amplitude $\sim$20 mas on a relatively short
timescale, but the period is highly uncertain.  We see evidence of a
stellar cycle with a period of about a decade in the 11 years of data
in-hand, as shown in the left panel of Figure \ref{fig:var_weirdos}.

\hspace{-22pt}{\bf SCR 0723-8015AB (0723-8015), Figure
  \ref{fig:perturbations1}:} This star exhibits a large perturbation
in both axes, but the orbit has not yet wrapped, even after 13 years
of data.  The star has a trigonometric distance (15.9 $\pm$ 0.7 pc)
that agrees with its photometric distance (17.1 $\pm$ 3.2 pc),
suggesting that the companion may be a white dwarf that contributes
minimal light at $I$.


\hspace{-22pt}{\bf SCR 0838-5855AB (0838-5855), Figure
  \ref{fig:perturbations1}:} This star was initially thought to be
within 10 pc, with a photometric distance of 8.0 $\pm$ 1.3 pc;
however, its trigonometric distance of 10.6 $\pm$ 0.1 pc places it
just beyond the 10 pc horizon.  This star shows evidence of a
perturbation in both axes in a dataset spanning 10 years, with an
amplitude in RA that is, so far, larger than that in DEC.


  

  
  


\hspace{-22pt}{\bf LP 788-1AB (0931-1717), Figure
  \ref{fig:perturbations2}:} This system exhibits a perturbation that
is evident in data acquired regularly since December 2011.  The
distances agree within the uncertainties: 12.5 $\pm$ 2 pc for its
photometric distance, compared to 14.7 $\pm$ 0.3 pc for its
trigonometric distance, implying a companion that emits little flux at
$I$.


\hspace{-22pt}{\bf LP 848-50AB (1042-2416), Figure
  \ref{fig:perturbations2}:} This object was initially thought to be
within 10 pc, based on its photometric distance of 9.7 $\pm$ 1.6 pc,
but the astrometric analysis shows it to be a close binary, with a
trigonometric distance of 11.3 $\pm$ 0.3 pc.  The orbit shows no signs
of wrapping in the current data set that spans six years.  It is
possible that the companion is a white dwarf, given the large
amplitude of the perturbation in the $R-$band --- note the 100 mas
scale in the panels of Figure \ref{fig:perturbations2} --- and
reasonable agreement of the two distances.

\hspace{-22pt}{\bf LTT 4004AB (1054$-$0718):} We find a new close
companion, B, indicated by elongated images but no resolution of the
two stars down to 1\farcs0.  There may be more than two stars present,
given that the trigonometric distance of 21.2 pc is well over 1.4
times that of the photometric distance of 12.2 pc.  Blinking
SuperCOSMOS images using Aladin reveals no other nearby stars that
could be background sources at the epochs of the astrometry
images. Alternately, this may be a young stellar binary.


\hspace{-22pt}{\bf LP 908-10 (1203$-$2923):} This object moved in
front of a background source that corrupted both the astrometry and
variability analyses.  Frames taken 2003$-$2013 were not used in the
reduction, and we do not report the variability here.


\hspace{-22pt}{\bf SCR 1230-3411AB (1230$-$3411), Figure
  \ref{fig:perturbations2}, \ref{fig:orbits}:} A perturbation is
evident in both the RA and DEC axes with a period of $\sim$5 years.
The trigonometric distance is 18.7 $\pm$ 0.6 pc versus the photometric
distance of 11.7 $\pm$ 1.8 pc, suggesting that the two stars have
similar luminosities in the $R-$band.  The luminosities are not
identical, however, or we would see no perturbation of the
photocenter.

\hspace{-22pt}{\bf L 327-121AB (1233$-$4826), Figure
  \ref{fig:perturbations2}:} A long-term perturbation is seen spanning
the full 5 years of the current data set.  The photometric distance
estimate is 10.3 $\pm$ 1.6 pc compared to the trigonometric distance
of 22.7 $\pm$ 0.7 pc, indicating that the unseen companion contributes
significant flux at $V$.  In fact, there may be more than one
companion to cause such discordant distances, or the system is young
and overluminous.

\hspace{-22pt}{\bf WT 1962AB (1259-0730), Figure
  \ref{fig:perturbations2}:} We detect a long-term astrometric
perturbation, likely caused by a low mass companion because minimal
light is contributed by the unseen secondary: the photometric distance
is 44.7 $\pm$ 7.2 pc and the trigonometric distance is consistent at
35.2 $\pm$ 1.7 pc.


\hspace{-22pt}{\bf L 408-123AB (1545$-$4330), Figure
  \ref{fig:perturbations2}, \ref{fig:orbits}:} The photometric
distance of 15.5 $\pm$ 2.4 pc is closer than the trigonometric
distance of 19.2 $\pm$ 0.4 pc.  A clear perturbation is seen in both
axes in frames taken at $R$ with an orbit that has wrapped, indicating
a period of $\sim$9 yr.



\hspace{-22pt}{\bf G 169-29 (1650$+$2226), Figure \ref{fig:var_weirdos}:}
As illustrated in the second panel of Figure \ref{fig:var_weirdos},
and as first reported in \citet{Hosey(2015)}, this object shows a
remarkable increase in brightness of $\sim$60 mmag (6\%) that now
stretches to 15 years in $R$.



\hspace{-22pt}{\bf G 154-043AB (1803-1858), Figure
  \ref{fig:perturbations3}:} This system shows a perturbation in the
DEC axis only that is longer than our current coverage of five years.
The estimated photometric distance is 9.9 $\pm$ 1.6 pc, whereas the
trigonometric distance is 13.0 $\pm$ 0.2 pc, implying significant flux
from the unseen companion at $R$.

\hspace{-22pt}{\bf LTT 7246AB (1815-1924):} We find a new companion,
B, at separation of 3.43\arcsec~at $\theta$ $=$ 218.2$^{\circ}$ in
2011, and at 3.53\arcsec~at $\theta$ $=$ 218.9$^{\circ}$ in 2015.
With a proper motion of 409 mas yr$^{-1}$, the source is physical.
The $\pi_{trig}$ reported is only for the A component, as the
secondary is faint.  The two stars have $\Delta$$V$ $=$ 2.2 mag,
$\Delta$$R$ $=$ 2.0 mag, and $\Delta$$I$ $=$ 1.5 mag.


\hspace{-22pt}{\bf LTT 7434AB (1845-2855), Figure
  \ref{fig:perturbations3}:} This target shows a perturbation, but the
orbit does not yet appear to have wrapped in our data, even with 15
years of astrometric coverage (there is a gap in the data sequence
from 2004--2010).  Its photometric distance is 1.4 times nearer than
its trigonometric distance (11.6 $\pm$ 1.8 pc versus 16.4 $\pm$ 0.4
pc), implying equal luminosity components, but because we see the
perturbation, the components must have different luminosities, or
there are more than two stars in the system. We note that
\citet{Bonfils(2013)} report this as an SB2.

\hspace{-22pt}{\bf L 489-43 (1852-3730):} A contaminating background
source corrupts the variability analysis of this star.  This source
may also have corrupted the $VRI$ photometry, as the photometric
distance is underestimated at 11.1 $\pm$ 1.9 pc, compared to its
trigonometric distance of 15.7 $\pm$ 0.3 pc.  However, the source does
not affect the astrometry, which shows residuals to the parallax and
proper motion fits of less than 2 mas on both the RA and DEC axes.
Given that no unseen companion is evident in the astrometry, the
photometric distance estimate likely suffers from additional flux from
the background source.

\hspace{-22pt}{\bf LHS 5348AB (1927-2811):} In frames with exceptional
seeing, this star is obviously elongated due to a companion found at
$\sim$1\arcsec.  The system's trigonometric distance is 12.7 $\pm$ 0.2
pc, compared to its photometric distance of 11.6 $\pm$ 1.9 pc,
indicating that the companion contributes very little light to the
system at $I$.


\hspace{-22pt}{\bf LEHPM 2-1265AB (2033-4903), Figure
  \ref{fig:perturbations3}:} This system shows a perturbation with a
period of about 5 years, seen only in the DEC axis.  The system's
photometric and trigonometric distances are 13.3 $\pm$ 2.1 pc and 16.7
$\pm$ 0.4 pc, respectively, so the secondary contributes only a modest
amount of light, if any, to the system at $R$.


\hspace{-22pt}{\bf L 427-34AB (2149-4133):} This is a binary with a
projected separation of $\sim$3\arcsec, making reductions challenging.
We measure $\pi_{trig}$ for both components that are consistent within
the relatively large uncertainties of $\sim$3 mas.  We do not report
variability results for this system, as the small separation corrupts
the flux measurements, but we do estimate the $\Delta$$V$ $\sim$0.3
mag.  The photometric distance for the unresolved pair is 11.4 $\pm$
2.0 pc, compared to the trigonometric distance of A of 16.0 $\pm$ 0.8
pc and the trigonometric distance of B of 14.7 $\pm$ 0.7 pc.


\hspace{-22pt}{\bf LEHPM 1-4771AB (2230-5344), Figure
  \ref{fig:perturbations3}, \ref{fig:orbits}:} A large perturbation is
evident in RA, but is muted in DEC.  A comparison of the photometric
distance (12 $\pm$ 1.8 pc) and the trigonometric distance (15 $\pm$
0.3 pc) indicates that the companion likely contributes only a modest
amount of flux to the system in $R$.

\hspace{-22pt}{\bf L 645-74B (2238$-$2921):} We present $\pi_{trig}$
for only the B component of this system, as the A component is
saturated in most of the images. However, we know the A component to
also be an M dwarf, based on its $(V-K)$ color of 4.4 mag
\citep{Winters(2015)}, so we include this system as a nearby M dwarf
pair.

\hspace{-22pt}{\bf SCR 2303-4650 (2303$-$4650):} There may be an
unseen companion orbiting with a period of less than 2 years.  The
typical cadence of our observations is not conducive to such an
orbital period, so monthly observations will be required to confirm or
refute the possible companion. For now, we note this object as a
candidate binary.

{\hspace{-22pt}\bf LTT 9828AB (2359-4405), Figure
  \ref{fig:perturbations3}, \ref{fig:orbits}:} This system shows a
clear perturbation, resulting in a $\sim$11 year orbit for the pair
that has wrapped in our data.  Its photometric distance of 17.6 $\pm$
2.7 pc and its trigonometric distance of 15.7 $\pm$ 0.6 pc agree
within the uncertainties, so the companion contributes very little
flux at $V$.

\section{Multiple Systems}
\label{sec:multiples}

\subsection{New Multiples --- Sixteen Unseen Companions}
\label{subsec:new}

In addition to distance measurements, long-term astrometric monitoring
also sometimes reveals the presence of unseen companions\footnote{We
  note that {\it component} refers to any member of a multiple
  system. The {\it primary} is either a single star or the most
  massive (or brightest in the $V-$band) component in the system, and
  {\it companion} refers to a member of a multiple system that is less
  massive (or fainter, again in the $V-$band) than the primary star.}
at small separations ($<$1\arcsec) from the star being investigated.
Previous publications in this series (see references in \S
\ref{sec:intro}), as well as \citet{Bartlett(2009)} have presented
results on a few of these new binaries, which are valuable for mass
determinations due to their short orbital periods.  Sixteen of the
systems presented here were found to exhibit astrometric perturbations
indicative of unseen companions. The arcs of the orbits were fit and
removed, the target positions adjusted, and a new solution for the
parallax and proper motion was obtained. These perturbations,
illustrated in Figures \ref{fig:perturbations1},
\ref{fig:perturbations2}, and \ref{fig:perturbations3}, represent
changes in the positions of the photocenters of the target stars
evident after the parallax and proper motions have been removed. The
$\pi_{trig}$ values given in Table \ref{tab:ctiopi_data} for these 16
systems represent values that include adjustments for orbital motion
and are noted.

From \citet{vandeKamp(1975)}, the magnitude of the perturbation in the
photocenter, $\alpha$, follows the relation $\alpha$ = (B $-$
$\beta$)~$a$, where B is the fractional mass M$_B$/(M$_A$ $+$ M$_B$),
the relative flux $\beta$ is expressed as $(1 + 10^{0.4\Delta
  m})^{-1}$, and $a$ is the semi-major axis of the relative orbit of
the two components.  From this relation, it is evident that
equal-mass/equal-flux binaries would show no perturbation, i.e., the
photocenter does not move.  Therefore, a star showing a perturbation
has a companion that is unequal in both flux and mass\footnote{One
  exception is the rare case of a red dwarf/white dwarf pair in which
  the two stars may have equal fluxes or equal masses, but not both at
  a given wavelength.}, but due to the degeneracy between the scaling
of the photocentric and relative orbits and mass ratio/flux
difference, there is uncertainty in the nature of the companion.
However, inferences {\it can} be made using the position of a star
with an unseen companion on the HR diagram, such as shown in Figure
\ref{fig:hr} (see $\S$\ref{sec:population} below for further
discussion).  For example, a star elevated in the HR diagram has a
companion that contributes significant flux, whereas a star that is
not elevated has a lower luminosity secondary --- for these red
dwarfs, perhaps a white dwarf or brown dwarf.




The nightly mean astrometric residual plots in RA and DEC for the 16
stars exhibiting perturbations are shown in Figures
\ref{fig:perturbations1}, \ref{fig:perturbations2}, and
\ref{fig:perturbations3}.  A star without a perturbation would have
residuals very close to the `0' line, such as shown in the upper left
panels of Figure \ref{fig:perturbations1} for SCR 1656-2046.  The
magnitudes of most perturbations are typically less than $\sim$20 mas,
but several stars show much larger perturbations --- SCR 0723-8015, LP
848-050AB, L 408-123AB, LTT 7434AB, and LTT 9828AB.  Note that the
y-axis scales for these five systems extend to 60 mas or 100 mas,
rather than the standard 40 mas plotted for the other 12 stars.  We
note that there are gaps in the temporal coverage for some stars, as
they were removed but then re-added to the astrometry program after a
period of a few years. Systems within 25 pc will continue to be
monitored to complete orbital coverage and improve orbital elements.


\subsection{New Multiples --- Orbital Fits for Four Systems}
\label{subsubsec:orbits}


For four of the unresolved systems, we see clear perturbations in both
the RA and DEC axes for which the astrometry coverage spans what
appear to be full orbits.\footnote{LP 349-25AB also has a wrapped
  orbit in our data, but we have HST-WFC3 data that will provide a
  detailed picture of the system and which will be described in a
  future paper in this series.} For the other 12 systems exhibiting
perturbations, the orbits have not yet wrapped due to insufficient
time coverage; thus, we do not provide the orbital elements here, as
they are not useful in isolation.

We use the methodology described in \citet{Hartkopf(1989)} that uses a
least-squares technique to solve for the orbital elements, given
initial guesses of three elements: orbital period ({\it P}), epoch of
periastron ({\it T}), and eccentricity ({\it e}).  Sets of initial
guesses are typically used to explore possible combinations for
orbital parameters to find the true minimum for the fit, rather than a
local minimum. The orbital fits for the four systems are shown in
Figure \ref{fig:orbits} and the elements for the photocentric orbits
are given in Table \ref{tab:orbits}.  The periods and semimajor axes
are generally reliable, while the other orbital elements are less
well-determined.  We also include the filter in which the observations
were made because the brightness ratios between components and the
consequent photocenter perturbations will vary with wavelength.



\voffset0pt{
\begin{deluxetable}{lcccccccc}
\tabletypesize{\tiny}
\tablecaption{Orbital Parameters}
\setlength{\tabcolsep}{0.03in}
\tablewidth{0pt}
\tablehead{\colhead{Name}                        &
           \colhead{filter}                      &
           \colhead{$P$}                         &
           \colhead{$a$}                         &
           \colhead{$e$}                         &
           \colhead{$i$}                         &
           \colhead{Long. Nodes ($\Omega$)}      &
           \colhead{Long. Periastron ($\omega$)} &
           \colhead{$T$}                         \\

	   \colhead{   }                         &
	   \colhead{   }                         &
           \colhead{(yr)}                        &
           \colhead{(mas)}                       &
           \colhead{}                            &
           \colhead{(deg)}                       &
           \colhead{(deg)}                       &
	   \colhead{(deg)}                       &
           \colhead{(yr)}                        \\
                                                 
           \colhead{(1)}                         &
           \colhead{(2)}                         &
           \colhead{(3)}                         &
           \colhead{(4)}                         &
           \colhead{(5)}                         &
           \colhead{(6)}                         &
           \colhead{(7)}                         &
           \colhead{(8)}                         &
           \colhead{(9)}                         }

\startdata
SCR 1230-3411AB & $R$ &  4.68 $\pm$ 0.08 & 20.1 $\pm$ 2.1 & 0.57 $\pm$ 0.17 &  86.2 $\pm$ 2.9 & 234.8 $\pm$  5.3 &  77.4 $\pm$   3.6 & 2011.36 $\pm$ 0.06 \\
L 408-123AB     & $R$ &  9.55 $\pm$ 0.22 & 31.2 $\pm$ 1.8 & 0.03 $\pm$ 0.07 &  91.9 $\pm$ 3.2 & 155.4 $\pm$  3.2 & 205.5 $\pm$ 137.4 & 2013.99 $\pm$ 3.61 \\ 
LEHPM 1-4771AB  & $R$ &  5.75 $\pm$ 0.13 & 20.2 $\pm$ 1.5 & 0.04 $\pm$ 0.06 & 106.5 $\pm$ 3.8 & 267.7 $\pm$  4.7 & 277.4 $\pm$  42.1 & 2012.26 $\pm$ 0.68 \\
LTT 9828AB      & $V$ & 11.39 $\pm$ 0.18 & 33.4 $\pm$ 1.8 & 0.32 $\pm$ 0.03 &  31.7 $\pm$ 7.7 & 133.6 $\pm$ 14.4 &  96.9 $\pm$  14.1 & 2001.74 $\pm$ 0.20 \\
\enddata

\label{tab:orbits}
\end{deluxetable}
}


\subsection{Multiplicity Fraction}
\label{subsec:known}

In addition to the 16 stars with unseen companions described in
$\S$5.1, there are 16 target stars with previously known companions,
or new companions revealed directly for the first time in our images
(e.g., LTT 4004B, LHS 5348B, etc.).  All 32 of the systems discussed
are currently known to be binaries, some of which have been presented
before in this series of papers; no higher order multiples, e.g.,
triples, quadruples, etc., are yet known. Table \ref{tab:multinfo}
lists the multiplicity information for all 32 systems in this sample
of 151 systems.  Listed are the separations ($\rho$) and position
angles ($\theta$) of the companions from their primaries, the dates of
the measurements, the techniques used to identify the binaries, and
the references for the separation measurements and the detection
technique.  Where available, the magnitude differences between the
components ($\Delta$mag), the filters in which the measurements were
made, and the references for the $\Delta$mags are also listed.


Newly presented separations were calculated using {\it SExtractor}
\citep{Bertin(1996)}. The delta-magnitudes ($\Delta$mag) for binaries
with small separations ($\lesssim$ 5\arcsec) were deblended via PSF
photometry, as described in $\S$\ref{subsubsec:vriphot}. The images
used for the $\Delta$mag calculations are usually those in which the
parallax was measured; thus, the filters listed are the parallax
filters for each system.


Of the 124 systems presented here that are within 25 pc, 29 are
binary, resulting in a multiplicity fraction of 23 $\pm$ 4\%.  We
consider this a lower limit, as there are likely additional multiples
yet to be identified, as described in the next section. Nonetheless,
the derived multiplicity fraction is in agreement with the recently
determined value found in the much larger sample of the first author
\citep{Wintersphd}, in which a value of 27.4 $\pm$ 1.3\% was measured
for a large sample of 1122 carefully vetted M dwarfs known to be
within 25 pc.


\begin{deluxetable}{lcccccccccc}
\tabletypesize{\tiny}
\tablecaption{Information for Confirmed Multiples}
\tablehead{\colhead{Name}                &
	   \colhead{RA}                  &
	   \colhead{DEC}                 &
           \colhead{$\rho$}              &
           \colhead{$\theta$}            &
           \colhead{Year}                &
           \colhead{Technique}           &
           \colhead{Ref}                 &
           \colhead{$\Delta$mag}         &
           \colhead{Filter}              &
           \colhead{Ref}                 \\

	   \colhead{   }                 &
           \colhead{(hh:mm:ss)}          &
           \colhead{(dd:mm:ss)}          &
           \colhead{(\arcsec)}           &
           \colhead{(deg)}               &
           \colhead{}                    &
           \colhead{}                    &
           \colhead{}                    &
	   \colhead{(mag)}               &
           \colhead{}                    &
           \colhead{}                    }

\startdata
LP 349-25AB          & 00 27 56.0  & $+$22 19 33  &   $<$1       & ...  &  2015  &  astdet &    1   &   0.38 &   H    &   3      \\ 
L 294-92AB           & 01 47 42.6  & $-$48 36 06  &   $<$1       & ...  &  2016  &  astdet &    1   &   ...  &  ...   &  ...     \\  
L 583-33AB           & 02 02 17.5  & $-$26 33 51  &      3.6     & 060  &  2001  &  visdet &    5   &   0.1  &$V_J$   &   1      \\ 
L 225-57AB           & 02 34 21.2  & $-$53 05 37  &   $<$1       & ...  &   ...  &    SB2  &    2   &   ...  &  ...   &  ...     \\
LP 831-45AB          & 03 14 18.0  & $-$23 09 31  &   $<$1       & ...  &  2016  &  astdet &    1   &   ...  &  ...   &  ...     \\
L 591-42AB           & 04 36 40.9  & $-$27 21 18  &     49.58    & 196  &  2001  &  visdet &    5   &   0.31 &$V_J$   &   5      \\ 
SCR 0644-4223AB      & 06 44 32.1  & $-$42 23 45  &      1.6     & 266  &  2009  &  visdet &    1   &   0.9  &$I_{KC}$&   1      \\ 
LP 382-56AB          & 06 57 11.7  & $-$43 24 51  &      2.2     & 032  &  2012  &  visdet &    1   &   0.1  &$V_J$   &   1      \\ 
SCR 0723-8015AB      & 07 24 00.8  & $-$80 15 22  &   $<$1       & ...  &  2016  &  astdet &    1   &   ...  &  ...   &  ...     \\
SCR 0838-5855AB      & 08 38 02.3  & $-$58 55 57  &   $<$1       & ...  &  2016  &  astdet &    1   &   ...  &  ...   &  ...     \\
LP 788-1AB           & 09 31 22.3  & $-$17 17 43  &   $<$1       & ...  &  2016  &  astdet &    1   &   ...  &  ...   &  ...     \\   
L 392-39AB           & 10 19 51.0  & $-$41 48 48  &     30.91    & 120  &  2001  &  visdet &    5   &   1.53 &$V_J$   &   5      \\ 
LP 848-50AB          & 10 42 41.3  & $-$24 16 04  &   $<$1       & ...  &  2016  &  astdet &    1   &   ...  &  ...   &  ...     \\
LTT 4004AB           & 10 54 42.0  & $-$07 18 33  &   $<$2       & ...  &  2016  &  visdet &    1   &   ...  &  ...   &  ...     \\
SCR 1230-3411AB      & 12 30 01.8  & $-$34 11 24  &      0.020   & ...  &  2016  &  astorb &    1   &   ...  &  ...   &  ...     \\
L 327-121AB          & 12 33 33.1  & $-$48 26 11  &   $<$1       & ...  &  2016  &  astdet &    1   &   ...  &  ...   &  ...     \\   
WT 1962AB            & 12 59 51.3  & $-$07 30 35  &   $<$1       & ...  &  2011  &  astdet &    1   &   ...  &  ...   &  ...     \\   
L 408-123AB          & 15 45 37.0  & $-$43 31 42  &      0.031   & ...  &  2015  &  astorb &    1   &   ...  &  ...   &  ...     \\ 
SCR 1630-3633AB      & 16 30 27.2  & $-$36 33 56  &      2.0     & 247  &  2015  &  visdet &    1   &   4.6  &$V_J$   &   1      \\
G 139-3AB            & 16 58 25.0  & $+$13 58 06  &      0.822   & 294  &  2008  &  spkdet &    4   &   0.48 & 754nm  &   4      \\
G 154-43AB           & 18 03 36.0  & $-$18 58 50  &   $<$1       & ...  &  2015  &  astdet &    1   &   ...  &  ...   &  ...     \\
L 43-72AB            & 18 11 15.3  & $-$78 59 23  &   $<$1       & ...  &  ....  &    SB2  &    2   &   ...  &  ...   &  ...     \\ 
LTT 7246AB           & 18 15 12.4  & $-$19 24 07  &      3.5     & 219  &  2015  &  visdet &    1   &   2.2  &$V_J$   &   1      \\ 
LTT 7419AB           & 18 43 12.5  & $-$33 22 46  &     14.90    & 354  &  2001  &  visdet &    5   &   5.99 &$V_J$   &   1      \\
LTT 7434AB           & 18 45 57.5  & $-$28 55 53  &   $<$1       & ...  &  2015  &  astdet &    1   &   ...  &  ...   &  ...     \\ 
L 345-91AB           & 18 48 41.4  & $-$46 47 08  &      2.3     & 032  &  2009  &  visdet &    1   &   0.1  &$V_J$   &   1      \\ 
LHS 5348AB           & 19 27 52.6  & $-$28 11 15  &   $<$2       & ...  &  2015  &  visdet &    1   &   ...  &  ...   &  ...     \\
LEHPM 2-1265AB       & 20 33 01.9  & $-$49 03 11  &   $<$1       & ...  &  2015  &  astdet &    1   &   ...  &  ...   &  ...     \\   
L 427-34AB           & 21 49 11.4  & $-$41 33 30  &      2.5     & 139  &  2011  &  visdet &    1   &   0.3  &$V_J$   &   1      \\ 
LEHPM 1-4771AB       & 22 30 09.5  & $-$53 44 56  &      0.020   & ...  &  2015  &  astorb &    1   &   ...  &  ...   &  ...     \\   
L 645-74AB           & 22 38 23.0  & $-$29 21 00  &     14.57    & 136  &  2002  &  visdet &    5   &   1.71 &$V_J$   &   5      \\ 
LTT 9828AB           & 23 59 44.8  & $-$44 05 00  &      0.033   & ...  &  2015  &  astorb &    1   &   ...  &  ...   &  ...     \\

\enddata

\tablerefs{(1) this work; (2) \citet{Bonfils(2013)}; (3)
  \citet{Forveille(2005)}; (4) \citet{Horch(2010)}; (5)
  \citet{Jao(2003)}; (6) \citet{Winters(2011)}. \label{tab:multinfo}}

\tablecomments{The codes for the techniques used to identify
  companions are as follows: {\bf astdet} --- astrometric detection of
  a perturbation of the photocenter, as discussed in $\S$5.1; {\bf
    astorb} --- astrometric orbit; {\bf SB2} --- double-lined
  spectroscopic binary; {\bf spkdet} --- detection using speckle
  interferometry; {\bf visdet} --- visual detection. The filters in
  which the $\Delta$mags were measured are either $V$ or $I$ on the
  Johnson-Kron-Cousins system, the filter at 754 nm used on the DSSI
  speckle instrument, or the 2MASS $H-$band.}

\end{deluxetable}
\clearpage



\subsection{Candidate Multiples or Young Stars?}
\label{subsec:candies}

As illustrated in Figure \ref{fig:ccd.trig}, there are 12 stars with
trigonometric distance measurements that are at least sqrt(2) times
larger than the estimated photometric distances.  These stars are
coincident with those elevated above the main sequence in the
observational HR diagram of Figure \ref{fig:hr}.  There are two
straightforward explanations for these stars --- they are either
unresolved multiples and/or young stars.

While we noted earlier that the sample had been vetted for young stars
prior to being measured for astrometry, the new parallax measurements
provide an extra parameter with which to analyze these stars for
indicators of youth. One attribute of young stars is a low tangential
velocity, $V_{tan}$, because as a group, young stars have not had the
many interactions with objects in the Galactic disk that lead to
disk-heating, and consequently higher $V_{tan}$ values (see
\citealt{Leggett(1992)} for red dwarfs in particular). While we cannot
pinpoint an individual young star by $V_{tan}$ alone, we can at least
examine the 12 outlier stars as a group.  $V_{tan}$ can be derived
from $\pi_{trig}$ and proper motion, and is listed in Table
\ref{tab:ctiopi_data} for the entire sample, and noted in Table
\ref{tab:cands} for the 12 stars that are overluminous.  As can be
seen in the histogram of Figure \ref{fig:vtan}, the $V_{tan}$ values
for the 12 candidates are not unusually low, which would be expected
for a sample of young stars.

Young stars are also generally coronally active, which results in
stars' detections in the X-ray and ultraviolet (UV) regions of the
electromagnetic spectrum. We searched for these 12 targets in the
ROentgen SATellite (ROSAT) All Sky Survey X-ray catalog
\citep{Voges(1999),Voges(2000)} and the Galaxy Evolution Explorer
(GALEX) DR5 UV catalog \citep{Bianchi(2011)}. Because our targets have
substantial proper motion, we used these proper motions to move the
targets to their 1991 positions before searching for them in ROSAT,
using a search radius of 25\arcsec, as per \citet{Voges(1999)}. The
midpoint of the GALEX survey was 2007, so we used the stars' proper
motions to find their 2007 positions before performing the search,
using a search radius of 5\arcsec.

Five of the 12 overluminous stars were found to have detections in the
ROSAT and GALEX catalogs, identifying them as coronally active, and
therefore likely young objects. One star LEHPM 2-3528 was detected in
ROSAT, while four stars were detected in the GALEX catalog: LTT 313,
SCR 0211-3504, SCR 2025-1534, and LP 567-63. LTT 313 had both FUV and
NUV detections, while the other four were detected in the NUV only. 



We also checked the space motions of the stars, to see if they
kinematically match any of the known nearby young moving groups. For
this, we used the LocAting Constituent mEmbers In Nearby Groups
(LACEwING) moving group identification code (Riedel et al., in
prep). LACEwING uses positions, proper motions, parallaxes, and radial
velocities to identify objects that might share the space motions and
positions of 13 nearby young moving groups and three open clusters
within 100 parsecs of the Sun. It has two modes, a field star mode in
which it assumes nothing is known about the age of the star, and a
young star mode, where it is assumed that the object is young. As
these 12 stars are overluminous, we used the young star mode for all
of them. Although we do not have radial velocities for these stars,
using our positions, proper motions, and parallaxes, we identify SCR
0027-0806 as a low-probability (28\%) member of the $\sim$45 Myr-old
\citep{Kraus(2014),Bell(2015)} Tucana-Horologium moving group; SCR
0724-3125 low-probability member of the $\sim$200 Myr old Carina-Near
moving group \citep{Zuckerman(2006)}, with a probability of 28\%; SCR
1901-3106 and SCR 2025-1534 as low- and moderate-probability members
of the $\sim$130-150 Myr old AB Doradus moving group
\citep{Luhman(2005),Bell(2015)} (probabilities of 27\% and 57\%,
respectively); and LP 567-63 as a member of the $\sim$50 Myr old Argus
moving group \citep{Torres(2008)}, with a probability of 38\%. We note
that both SCR 2025-1534 and LP 567-63 were found to have detections in
GALEX.

Based on these analyses, we suspect that half of these highlighted
systems are unresolved multiples, with the other half being possibly
young.





\begin{deluxetable}{lcccc}
\centering
\tabletypesize{\small}
\tablecaption{Candidate Multiples \& Young Stars}
\setlength{\tabcolsep}{0.03in}
\tablewidth{0pt}
\tablehead{\colhead{Name}                &
	   \colhead{RA}                  &
	   \colhead{DEC}                 &
	   \colhead{$V_{tan}$}            &
	   \colhead{Youth}               \\

	   \colhead{   }                 &
           \colhead{(hh:mm:ss)}          &
           \colhead{(dd:mm:ss)}          &
           \colhead{(km s$^{-1}$)}        &
           \colhead{Indicator}           }

\startdata
SCR 0027-0806   & 00 27 45.4  &  $-$08 06 05  &  22.6  & Tuc-Hor  \\ 
LTT 313         & 00 35 38.1  &  $-$10 04 19  &  30.1  & GALEX    \\
L 2-60          & 01 29 20.8  &  $-$85 56 11  &  63.4  &          \\
SCR 0211-0354   & 02 11 51.7  &  $-$03 54 03  &  23.7  & GALEX    \\
LEHPM 2-3528    & 06 07 58.1  &  $-$61 15 11  &  42.1  & ROSAT    \\
SCR 0724-3125   & 07 24 21.2  &  $-$31 25 58  &  24.7  & Carina-Near  \\
SCR 0733-4406   & 07 33 42.7  &  $-$44 06 13  &  57.4  &          \\
L 532-12        & 08 54 02.4  &  $-$30 51 37  &  39.2  &          \\
LHS 5231        & 12 59 18.2  &  $-$00 10 33  &  64.8  &          \\
SCR 1901-3106   & 19 01 59.2  &  $-$31 06 45  &  43.0  & AB Dor    \\
SCR 2025-1534   & 20 25 08.6  &  $-$15 34 16  &  29.9  & AB Dor, GALEX    \\
LP 567-63       & 20 34 31.1  &  $-$32 31 00  &  26.1  & Argus, GALEX    \\

\enddata

\tablecomments{The youth indicators `GALEX' and `ROSAT' indicate
  detections in each respective catalog, while `Tuc-Hor', `AB Dor',
  `Carina-Near', and `Argus' indicate the young nearby moving groups
  to which the star in question has a probability of belonging, based
  on its kinematic information.}

\label{tab:cands}
\end{deluxetable}

\clearpage

\section{Discussion: The Nearby M Dwarf Population}
\label{sec:population}

Recent noteworthy surveys have considerably improved our knowledge of
the northern M dwarf population, with \citet{Dittmann(2014)}
contributing 1507 and \citet{Finch(2016)} adding 1059 $\pi_{trig}$
measurements for our nearby low-mass neighbors. Yet, the southern sky
has remained largely untouched. RECONS' work is improving the census
of M dwarfs with accurate distances in the southern hemisphere.


Prior to this paper, there were 448 southern M dwarf systems with
accurate (uncertainties less than 10 mas) published $\pi_{trig}$
placing them within 25 pc.  This number includes measurements from YPC
(192), HIP (88), previous publications in this series by RECONS (131),
and a few dozen systems from other efforts.  The trigonometric
distances for the 116 systems within 25 pc and south of DEC = 0
presented here increase the census of M dwarf systems in the southern
sky by 26\%. The population density diagram for the southern sky shown
in Figure \ref{fig:pop_den}, in which we use use $M_V$ as a proxy for
mass, provides a gauge of the uniformity of systems within a volume
stretching to 25 pc, broken into eight equal volume shells, with the
first shell incorporating half of the distance, but one-eighth of the
volume, to the horizon.  The 116 new southern members added by this
work are indicated in red.  Intrinsically faint, late-type M dwarfs
are more prominent at closer distances than their mid-type
counterparts at further distances.\footnote{All systems reported here
  are beyond 10 pc, as the closer systems will be described in an
  upcoming paper in this series.} The paucity of late M dwarfs at
larger distances is clearly an observational bias, as their faintness
has precluded their identification and parallax measurement to date.











\section{Future}
\label{sec:future}

Members of the RECONS group are stellar cartographers, driven to
discover and characterize the nearby population of stars. While
previous work has reported the census of the 5 pc sample
\citep{Henry(2015)} and 10 pc sample \citep[and in prep]{Henry(2006)},
we are expanding our investigations of the nearest stars to 25 pc.
This 25 pc horizon limit has been adopted as a match to the canonical
distance limit of the The Third Catalogue of Nearby Stars
\citep{Gliese(1991)} and ultimately should include $\sim$5000 stellar
systems.

Development by the RECONS group of a modern catalog listing all stars,
brown dwarfs, and planets located within 25 pc, with distances
determined only via $\pi_{trig}$, has been underway for the past few
years.  In this compendium, called the RECONS 25 Parsec Database, is a
wealth of information on each system, with only data from carefully
vetted publications selected for inclusion.  The current version, as
of 01 August 2016, includes 2826 systems containing 3801 stars, brown
dwarfs, and planets, and we anticipate releasing it to the
astronomical community in the near future.  The main criterion for
inclusion is that each system must have a published trigonometric
parallax of at least 40 mas, with an uncertainty $\leq$ 10 mas; thus,
124 of the systems reported here will be new entries to the database.


The long-anticipated results from {\it Gaia} will measure distances
for many nearby stars currently lacking accurate $\pi_{trig}$
measurements. But, ground-based astrometry will continue to be able to
push to the end of the stellar main sequence at $M_V$ $=$ 21.5
\citep{Dieterich(2014)} all the way to 25 pc, a realm unlikely to be
reached by {\it Gaia}.  Thus, there will continue to be a need for
ground-based astrometry efforts. As the RECONS observing program began
at the 0.9m as an NOAO Survey in 1999, the data series now stretches
to over a decade for hundreds of nearby southern M dwarfs, providing
long-term astrometry and photometry coverage that will likely not be
available from {\it Gaia}, nor any other project for the foreseeable
future. Thus, this unique temporal dataset coverage is valuable for
studies of these low-mass neighbors, especially those exhibiting
perturbations for which mass determinations will be possible. In
addition, RECONS' long-term monitoring of these potential exoplanet
host stars enables investigations of their long-term photometric
trends, such as variability and stellar cycles, that may affect life
on any planets around these low-mass neighbors.




\section{Acknowledgments}

This research was made possible by NSF grants AST 05-07711, AST
09-08402, and AST 14-12026.  We also thank the members of the SMARTS
Consortium, who have enabled the operations of the small telescopes at
CTIO since 2003, as well as the observer support at CTIO, specifically
Edgardo Cosgrove, Arturo Gomez, Manuel Hernandez, Alberto Miranda,
Mauricio Rojas, Hernan Tirado, and Joselino Vasquez.  We thank the
referee for a very thorough review of the paper that has allowed us to
improve it.

Data products from the Two Micron All Sky Survey, which is a joint
project of the University of Massachusetts and the Infrared Processing
and Analysis Center/California Institute of Technology, funded by the
National Aeronautics and Space Administration and the National Science
Foundation have been used extensively, as have the SIMBAD database and
the Aladin and Vizier interfaces, operated at CDS, Strasbourg, France.

JGW is currently supported by a grant from the John Templeton
Foundation. The opinions expressed here are those of the authors and
do not necessarily reflect the views of the John Templeton Foundation.


\bibliographystyle{apj}
\bibliography{ref}

\clearpage


\begin{figure}[ht]
\minipage{0.50\textwidth}
\centering
{\includegraphics[scale=0.25,angle=90]{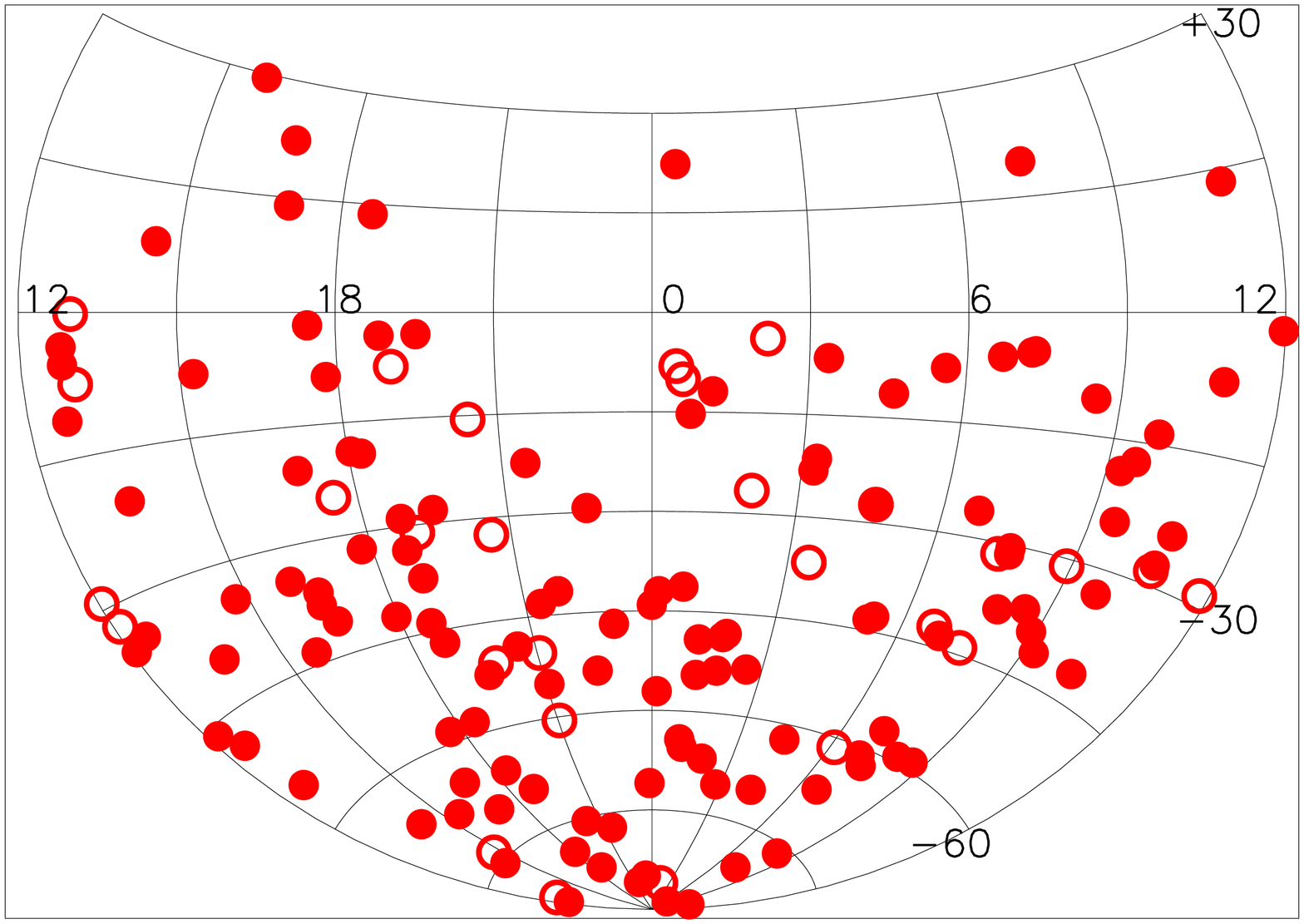}}
\endminipage\hfill
\minipage{0.50\textwidth}
\centering
{\includegraphics[scale=0.33,angle=90]{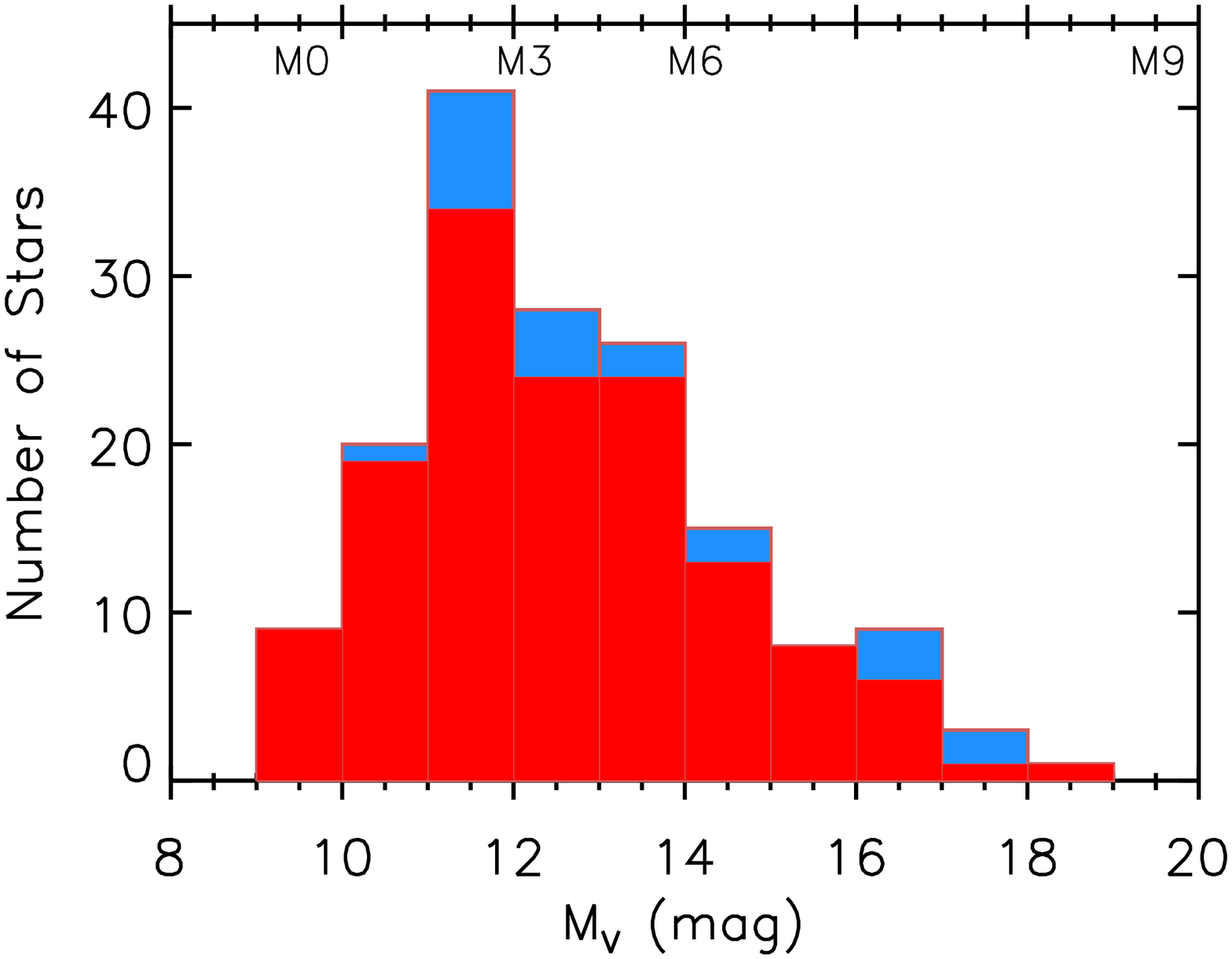}}
\endminipage\hfill
\vspace{15pt} \figcaption{(Left) Aitoff plot of the distribution of
  the sample on the sky. Solid points indicate the 131 stars (in 124
  systems) within 25 pc (note that eight are north of $\delta$ $=$
  0$^{\circ}$). Open points indicate the 30 objects (in 27 systems)
  with $\pi_{trig}$ placing them beyond 25 pc. (Right) The luminosity
  function for the stars presented here. In blue are indicated the 21
  unresolved multiples with confirmed close companions, where the
  $M_V$ has been calculated from blended photometry, while the 139
  objects with individual photometry are shown in red. Spectral type
  estimates are indicated along the top of the
  plot.  \label{fig:sample_def}}
\end{figure}



\begin{figure}
\includegraphics[scale=0.50,angle=90]{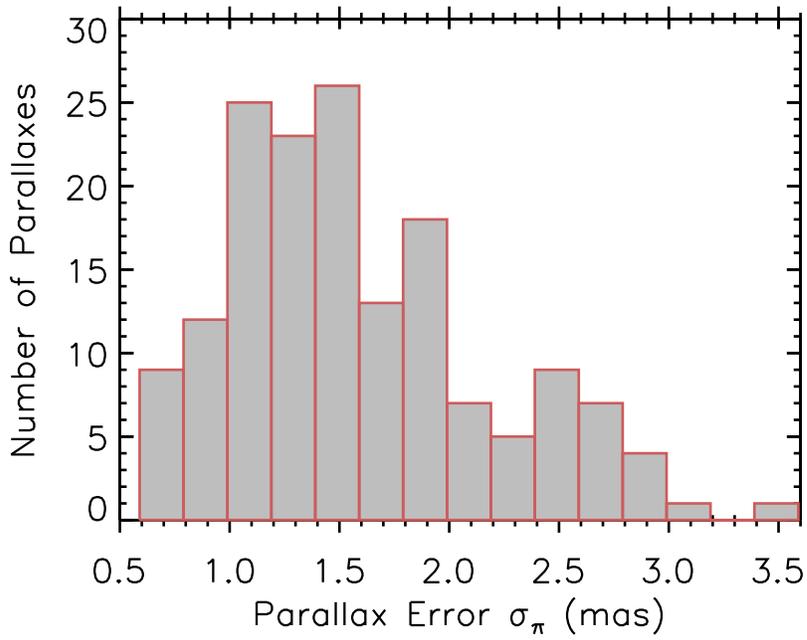} \figcaption{A
  histogram of the $\pi_{trig}$ uncertainties for the 161 parallaxes
  presented here, illustrating that most errors are less than 2 mas,
  with a mean parallax error for the entire sample of 1.6
  mas. \label{fig:pi_err}}
\end{figure} 


\begin{figure}
\includegraphics[scale=0.60,angle=90]{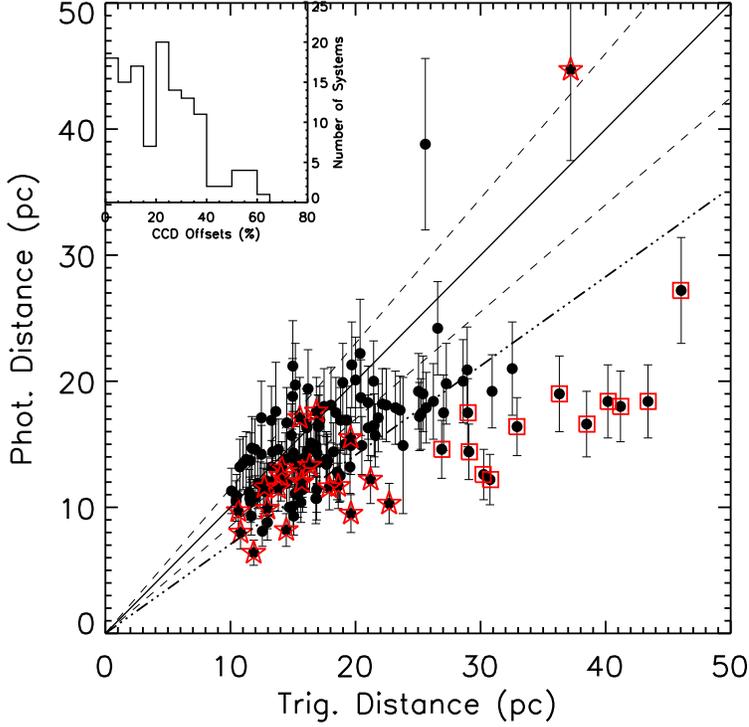} \figcaption{A
  comparison of distance estimates from $VRIJHK$ photometry
  vs.~distances measured using $\pi_{trig}$ for almost the entire
  sample of stars presented here. The six multiple systems with joint
  2MASS photometry that prevents the calculation of individual
  photometric distance estimates (SCR 0644-4223AB, LP 382-56AB, SCR
  1630-3633AB, LTT 7246AB, L 345-91AB, L 427-34AB) are not included in
  this plot, nor are the two stars with distances $>$ 50 pc (CE 440-87
  and WT 1928). Uncertainties on the trigonometric distances are
  smaller than the symbols. Known unresolved multiples with blended
  photometry are enclosed by open red stars. Candidate unresolved
  multiples are enclosed with open red squares. The diagonal solid
  line represents 1:1 correspondence in distances, while the dashed
  lines indicate the 15\% uncertainties associated with the CCD
  distance estimates.  The dash-dot line traces the location where the
  trigonometric distance exceeds the photometric estimate by a factor
  of sqrt(2), corresponding to an equal-luminosity pair of stars.  The
  inset histogram indicates the distribution of the distance offsets
  between the photometric and trigonometric distances for the single
  stars. For these single stars, the absolute mean offset is 21\%,
  slightly higher than the 15\% systematic error determined by
  \citet{Henry(2004)}. \label{fig:ccd.trig}}
\end{figure}


\begin{figure}
\includegraphics[scale=0.70,angle=90]{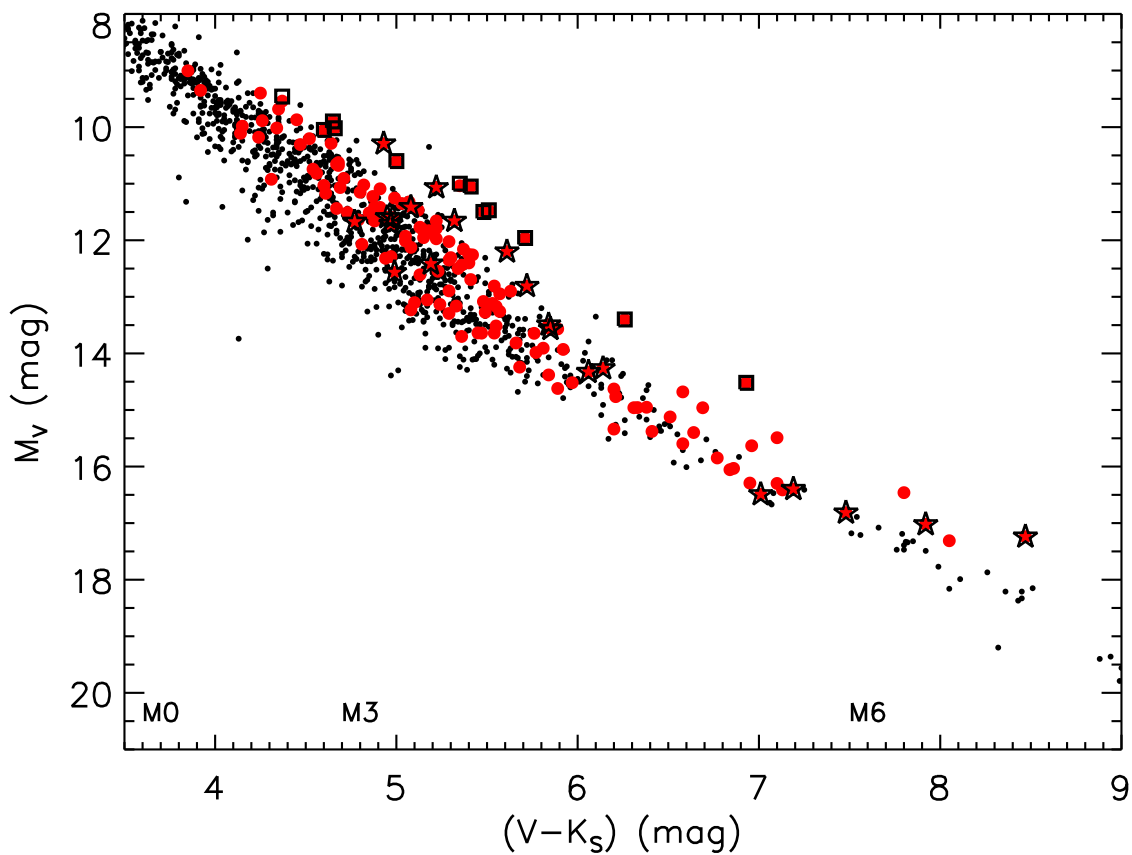} \figcaption{An
  observational HR diagram for the M dwarf systems presented here,
  with $M_{V}$ and ($V-K$) used as proxies for luminosity and
  temperature, respectively. Known multiples with blended photometry
  are enclosed in open stars. Candidate binaries are elevated above
  the main sequence or have distance mismatches and are enclosed in
  open squares. The small background points are presumed single stars
  from the RECONS 25 PC Database.  Spectral type estimates are given
  along the bottom; however, M9V corresponds to a $(V-K)$ color of 9.5
  and is thus off the plot. Not included on this plot are the six
  close multiple systems with joint 2MASS photometry.\label{fig:hr} }
\end{figure} 

\begin{figure} 
\includegraphics[scale=0.50,angle=90]{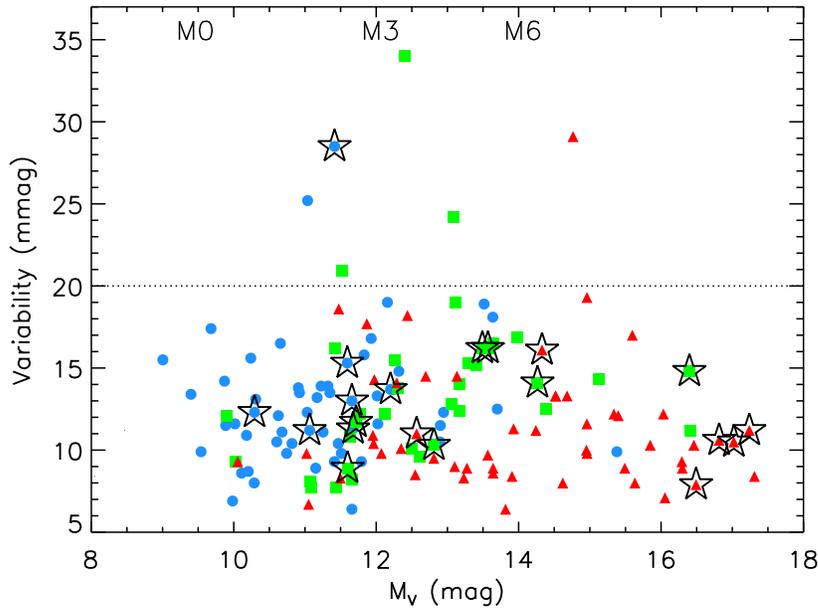}
\figcaption{Variability of the stars (in $VRI$) as a function of
  $M_V$. Stars observed in the $V-$band are indicated as blue points,
  in the $R-$band as green squares, and in the $I-$band as red
  triangles. Close binaries with blended photometry are noted with
  open black stars. The dotted line marks the 20 mmag (2\%)
  variability boundary between active (above) and inactive (below)
  stars. Spectral type estimates are indicated along the top of the
  plot; however, M9V corresponds to an $M_V$ magnitude of 20.0 and is
  thus off the plot.  \label{fig:var.vk}}
\end{figure}


\begin{figure}[ht]
\minipage{0.50\textwidth}
\centering
{\includegraphics[scale=0.33,angle=90]{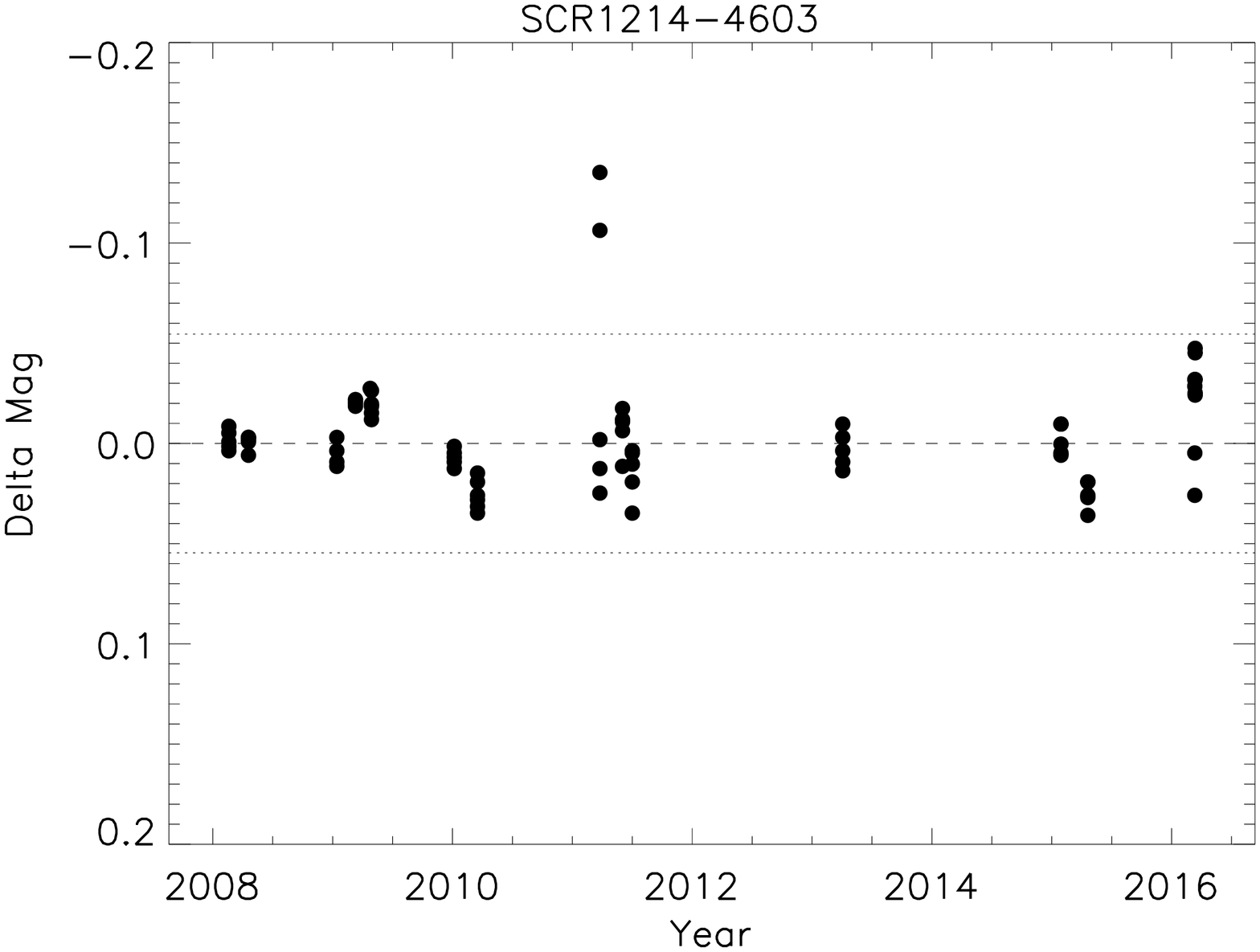}}
\endminipage\hfill
\minipage{0.50\textwidth}
\centering
{\includegraphics[scale=0.33,angle=90]{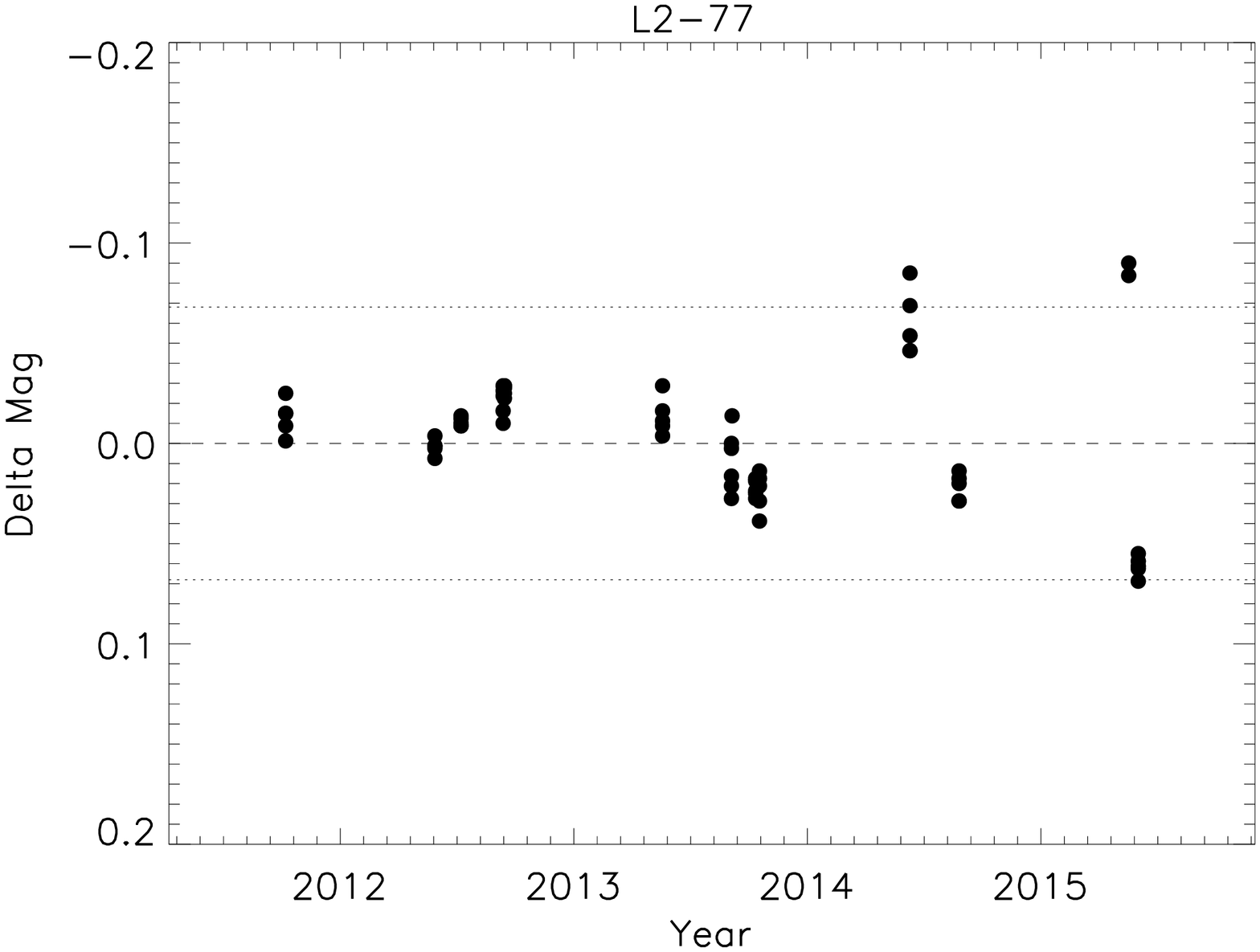}}
\endminipage\hfill
\vspace{15pt} \figcaption{(Left) SCR 1214-4603 showing evidence of a
  flare in 2011 with eight years of data. (Right) With almost four
  years of coverage, L 2-77 was fairly stable through 2012, but has
  recently started showing signs of flares and spots. Dashed lines in
  both panels are two times the standard deviations of the individual
  points. \label{fig:var_weirdos2}}
\end{figure}

\begin{figure}[ht]
\minipage{0.50\textwidth}
\centering
{\includegraphics[scale=0.33,angle=90]{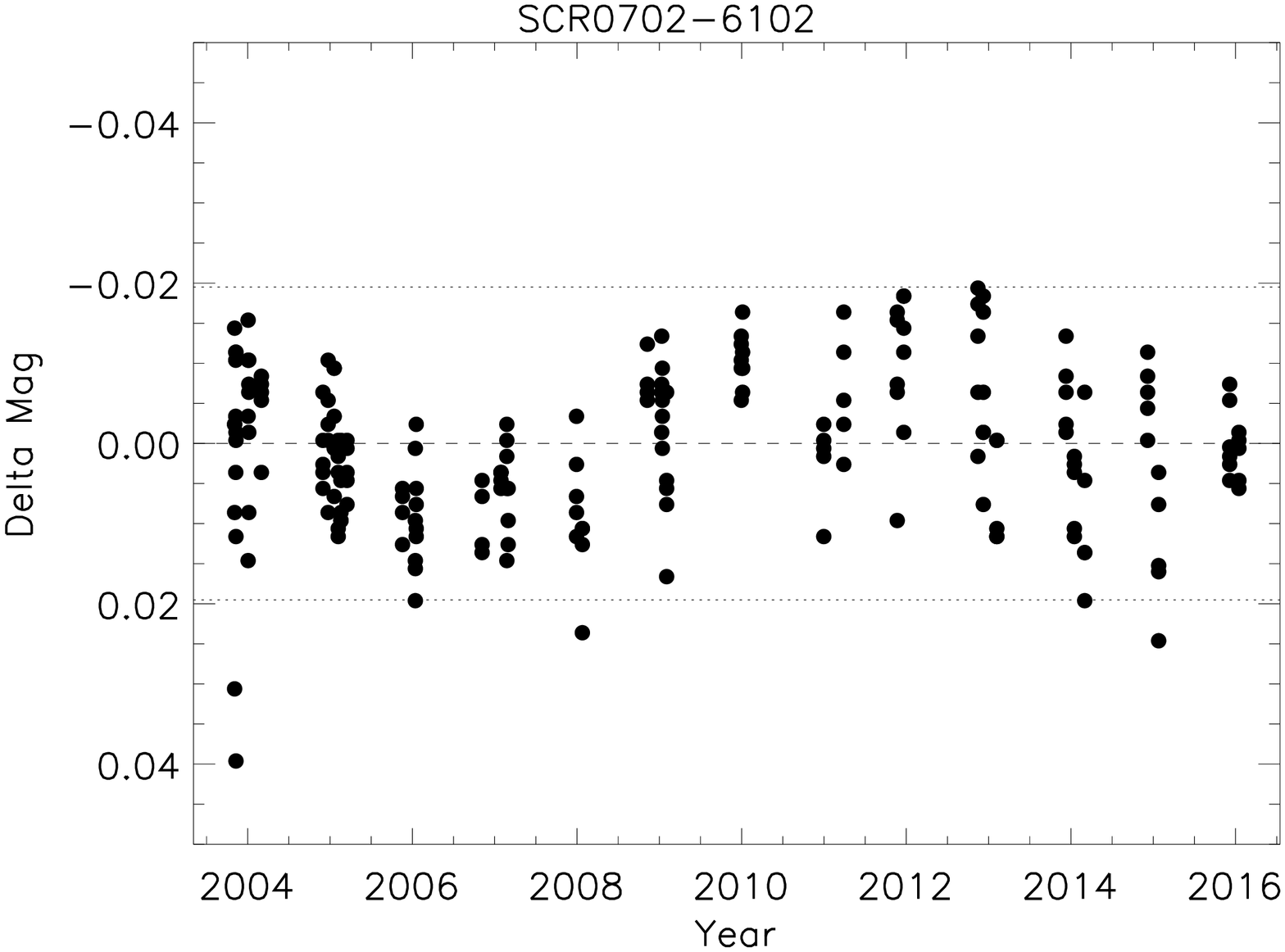}}
\endminipage\hfill
\minipage{0.50\textwidth}
\centering
{\includegraphics[scale=0.33,angle=90]{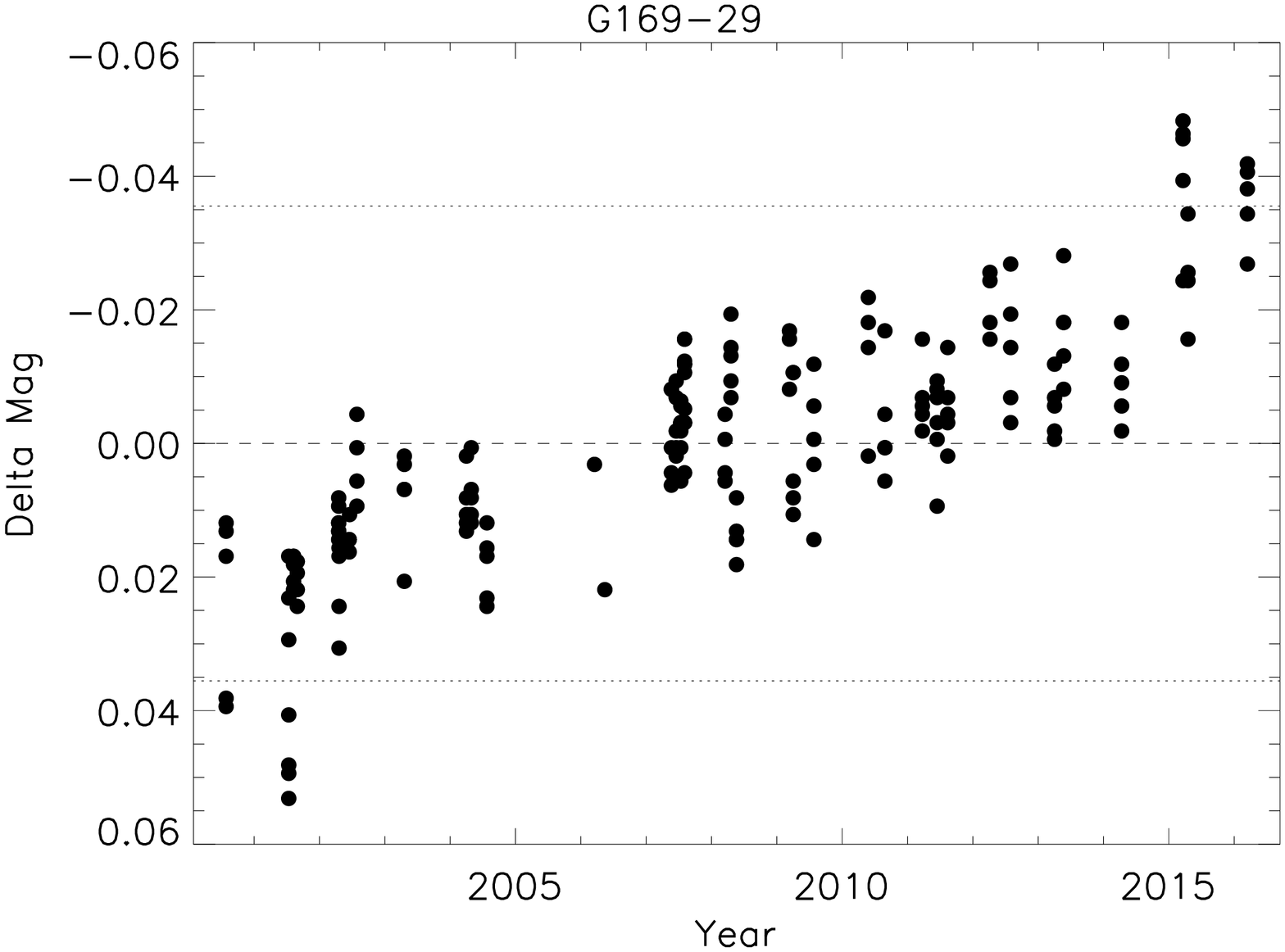}}
\endminipage\hfill
\vspace{15pt} \figcaption{(Left) Clear stellar cycle of SCR 0702-6102
  with a period on the order of a decade. (Right) The trend of
  increasing brightness of G 169-29 highlighted in \citet{Hosey(2015)}
  now extends to 15 years.  Dashed lines in both panels are two times
  the standard deviations of the individual
  points. \label{fig:var_weirdos}}
\end{figure}

\begin{figure}[ht]
\minipage{0.50\textwidth}
\centering
{\includegraphics[scale=0.33,angle=90]{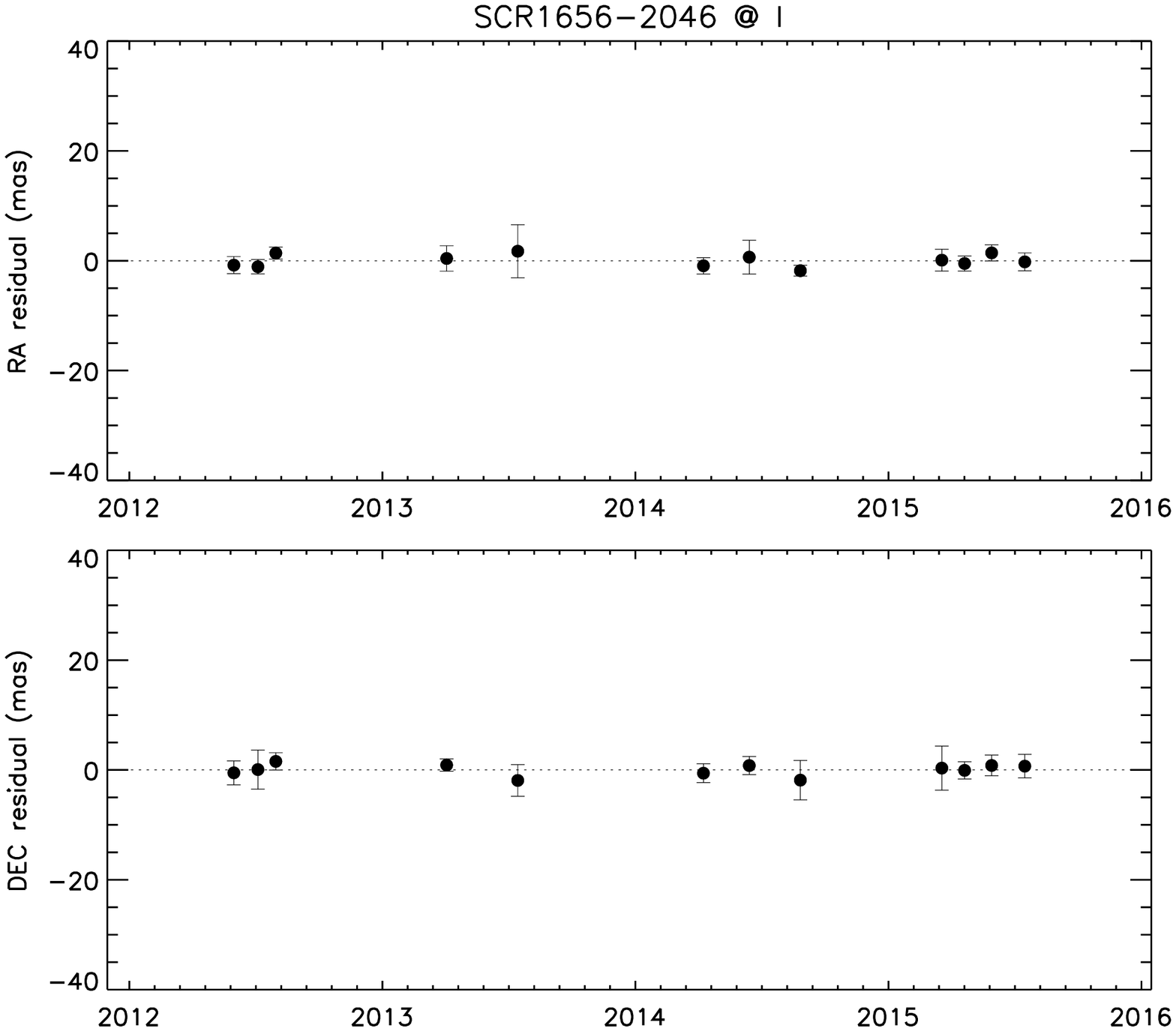}}
\endminipage\hfill
\minipage{0.50\textwidth}
\centering
{\includegraphics[scale=0.33,angle=90]{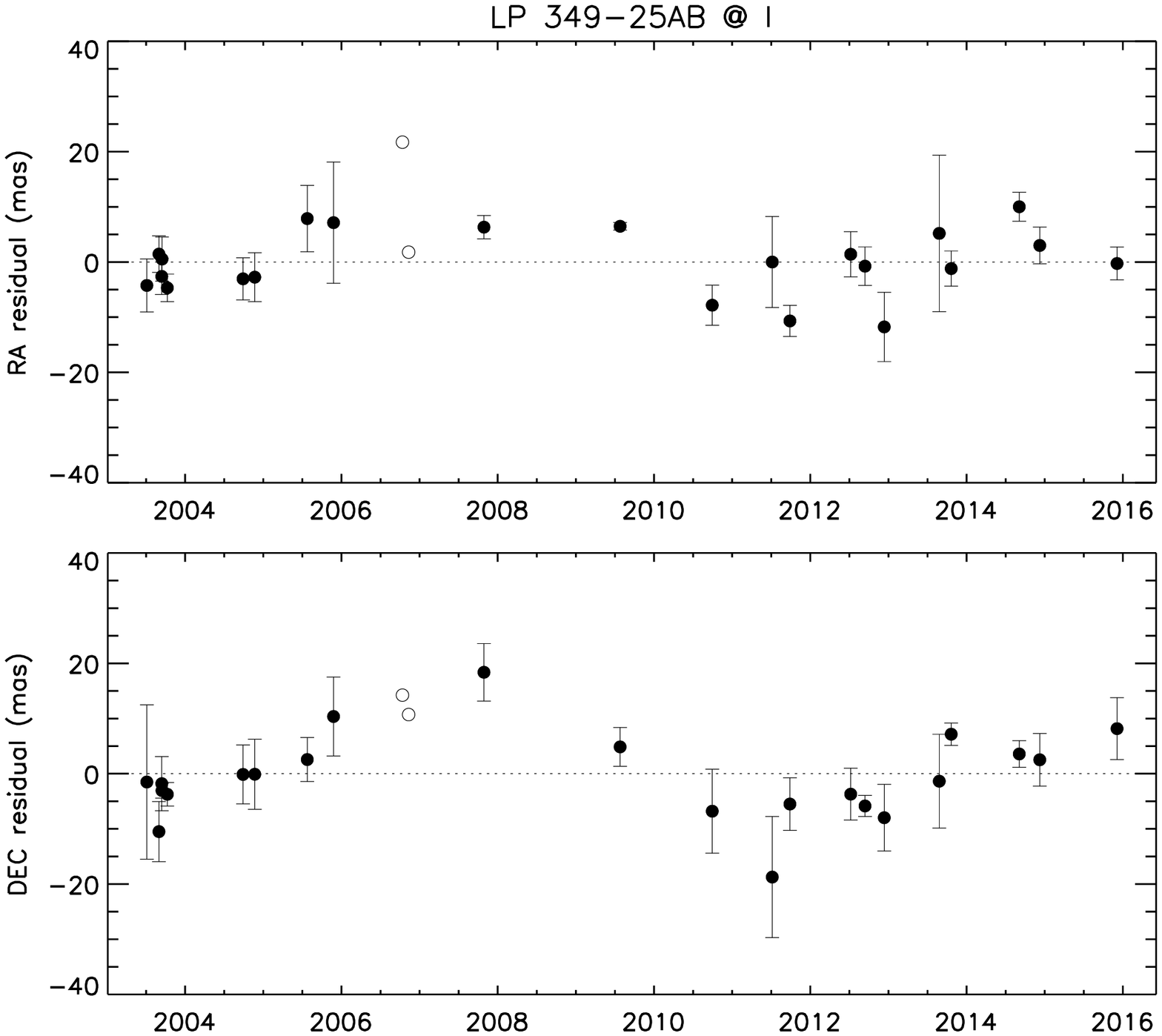}}
\endminipage\hfill
\vspace{15pt}
\minipage{0.50\textwidth}
\centering
{\includegraphics[scale=0.33,angle=90]{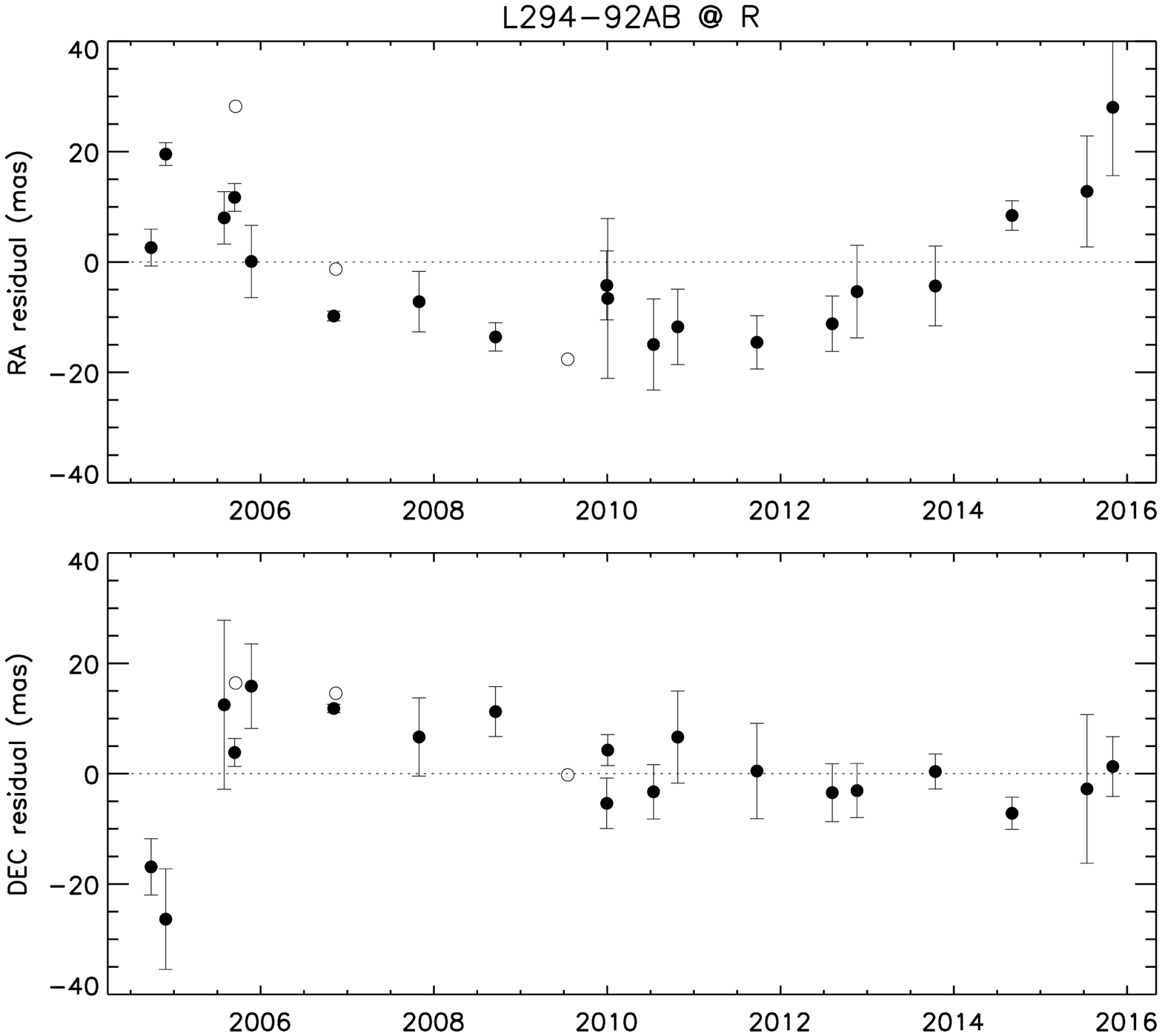}}
\endminipage\hfill
\minipage{0.50\textwidth}
\centering
{\includegraphics[scale=0.33,angle=90]{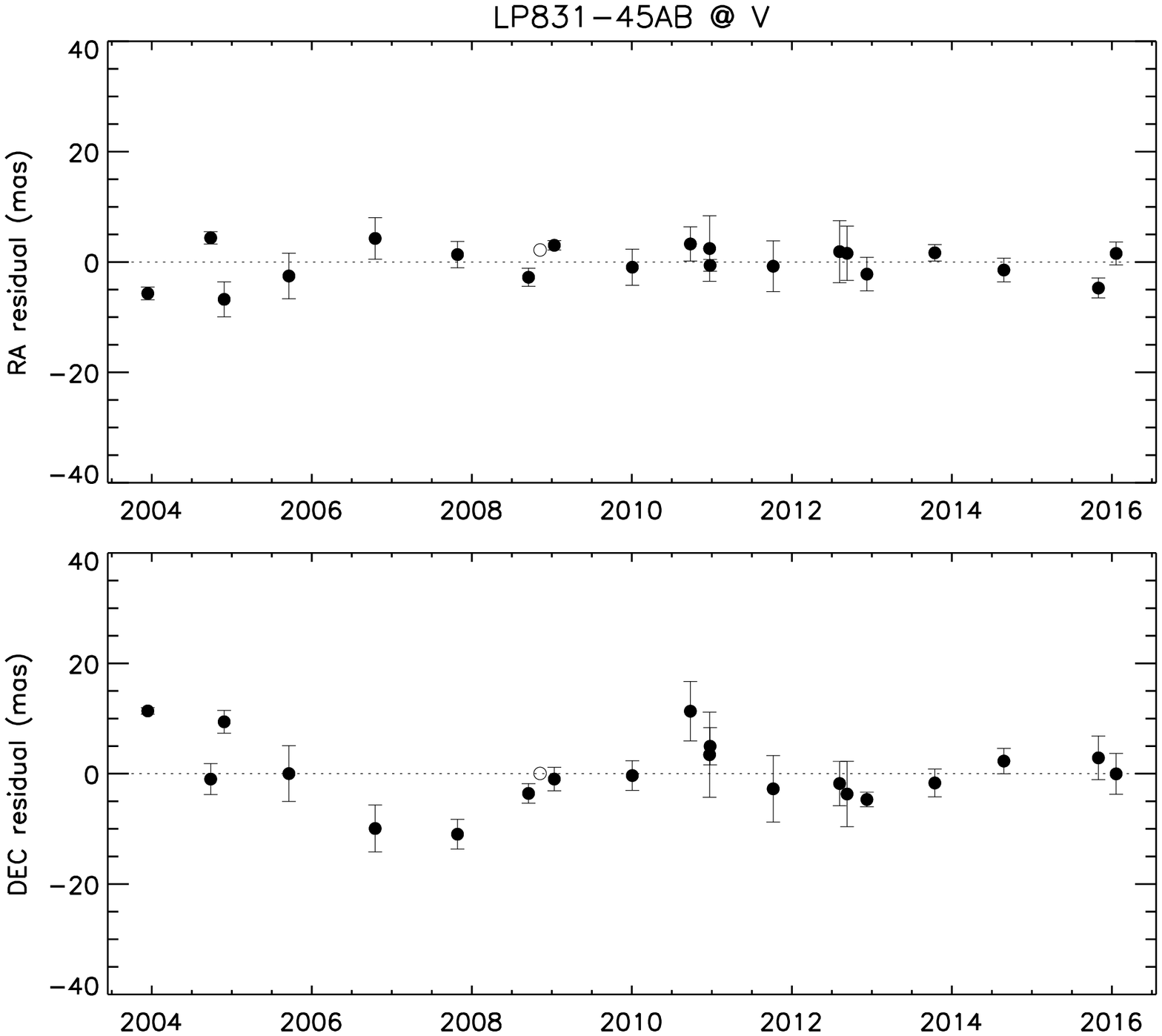}}
\endminipage\hfill
\vspace{15pt}
\minipage{0.50\textwidth}
\centering
{\includegraphics[scale=0.33,angle=90]{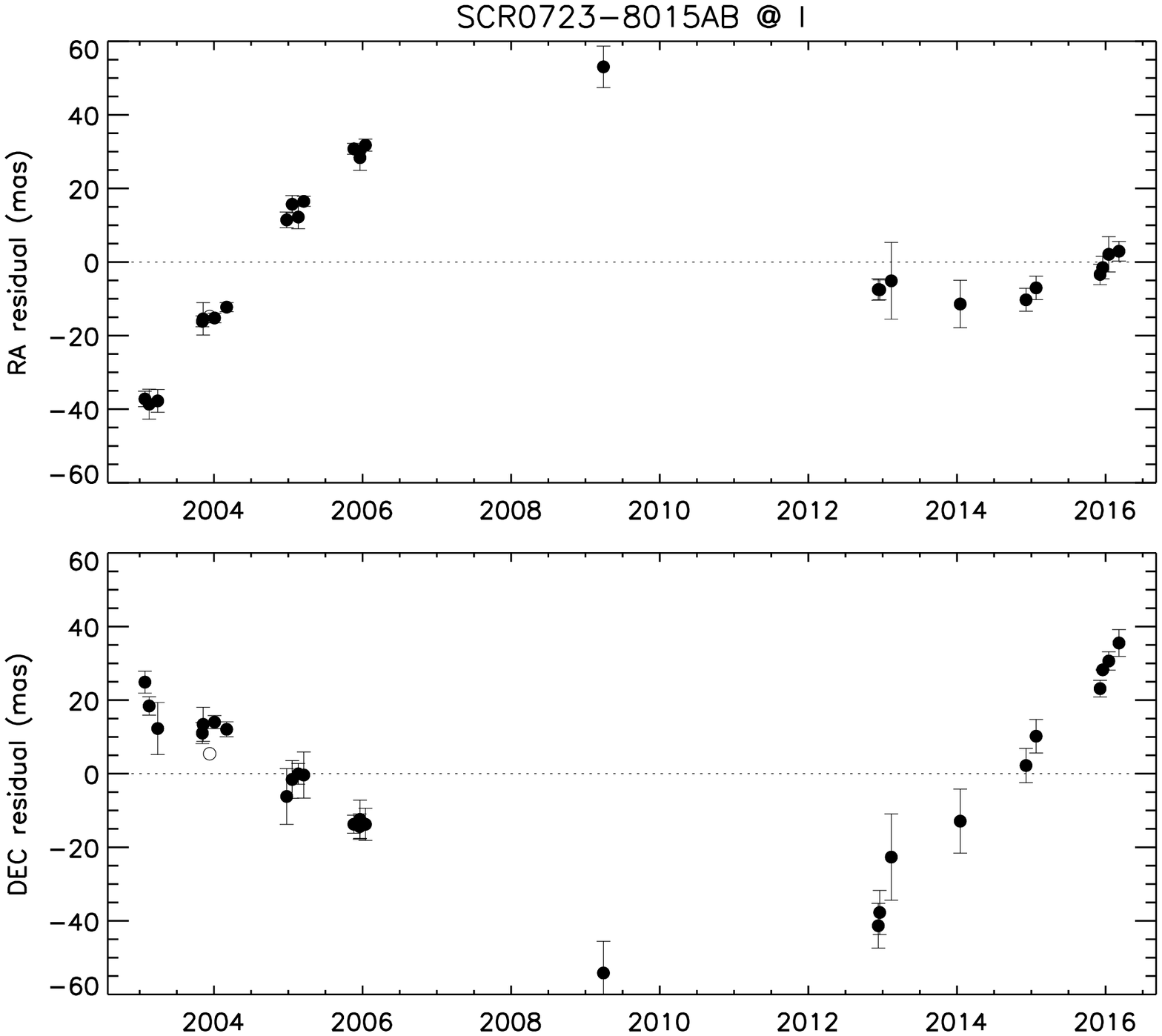}}
\endminipage\hfill
\minipage{0.50\textwidth}
\centering
{\includegraphics[scale=0.33,angle=90]{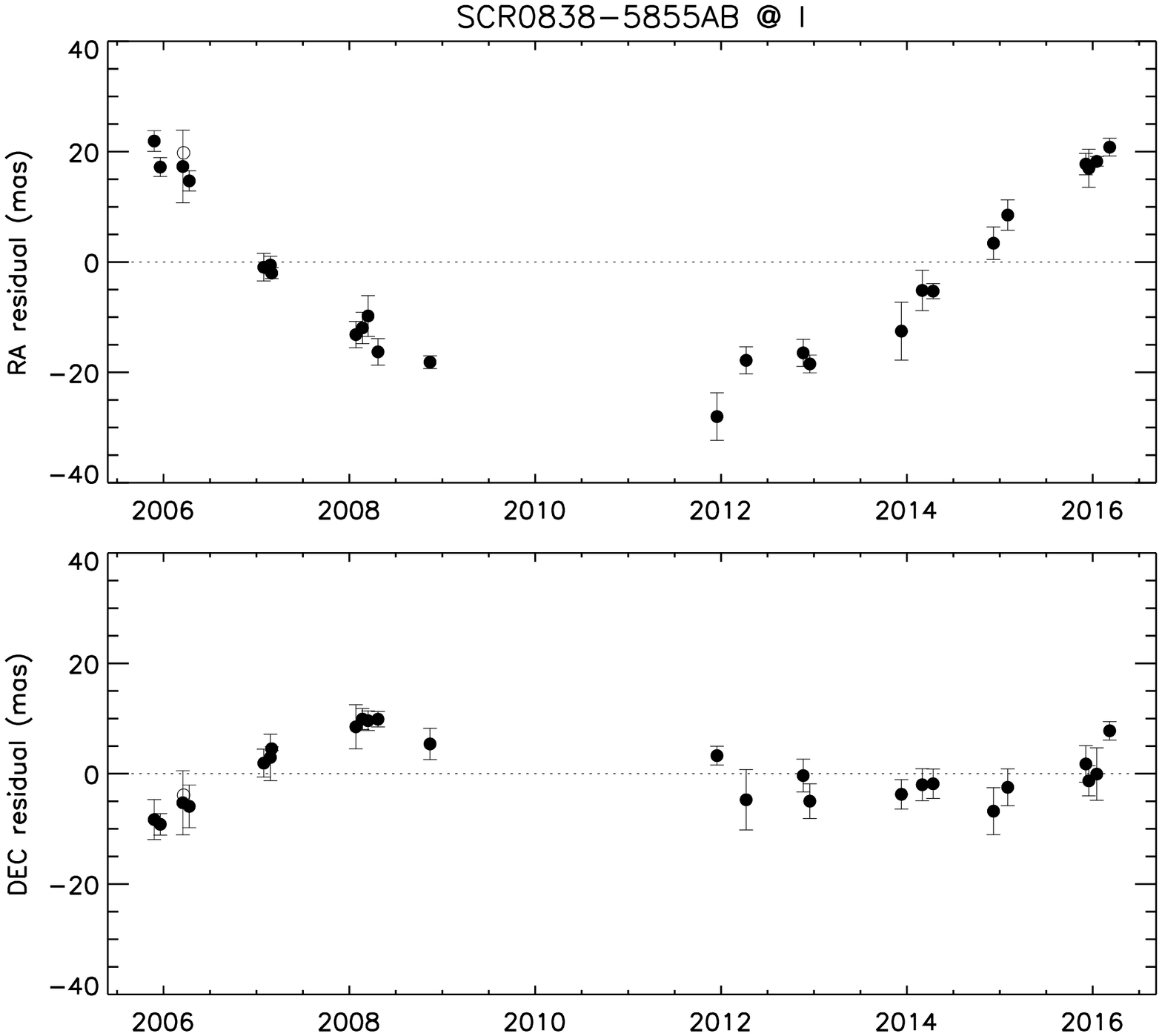}}
\endminipage\hfill
\vspace{10pt}
\caption{Nightly mean astrometric residual plots in RA and DEC for SCR
  1656-2046, LP 349-25AB, L 294-92AB, LP 831-45AB, SCR 0723-8015AB,
  and SCR 0838-5855AB. The astrometric signatures of each system's
  proper motion and parallax have been removed. SCR 1656-2046 is shown
  as an example of a single (i.e., {\it non-perturbed}) star with
  residuals clustered closely around zero. \label{fig:perturbations1}}
\end{figure}


\begin{figure}[ht]
\minipage{0.50\textwidth}
\centering
{\includegraphics[scale=0.33,angle=90]{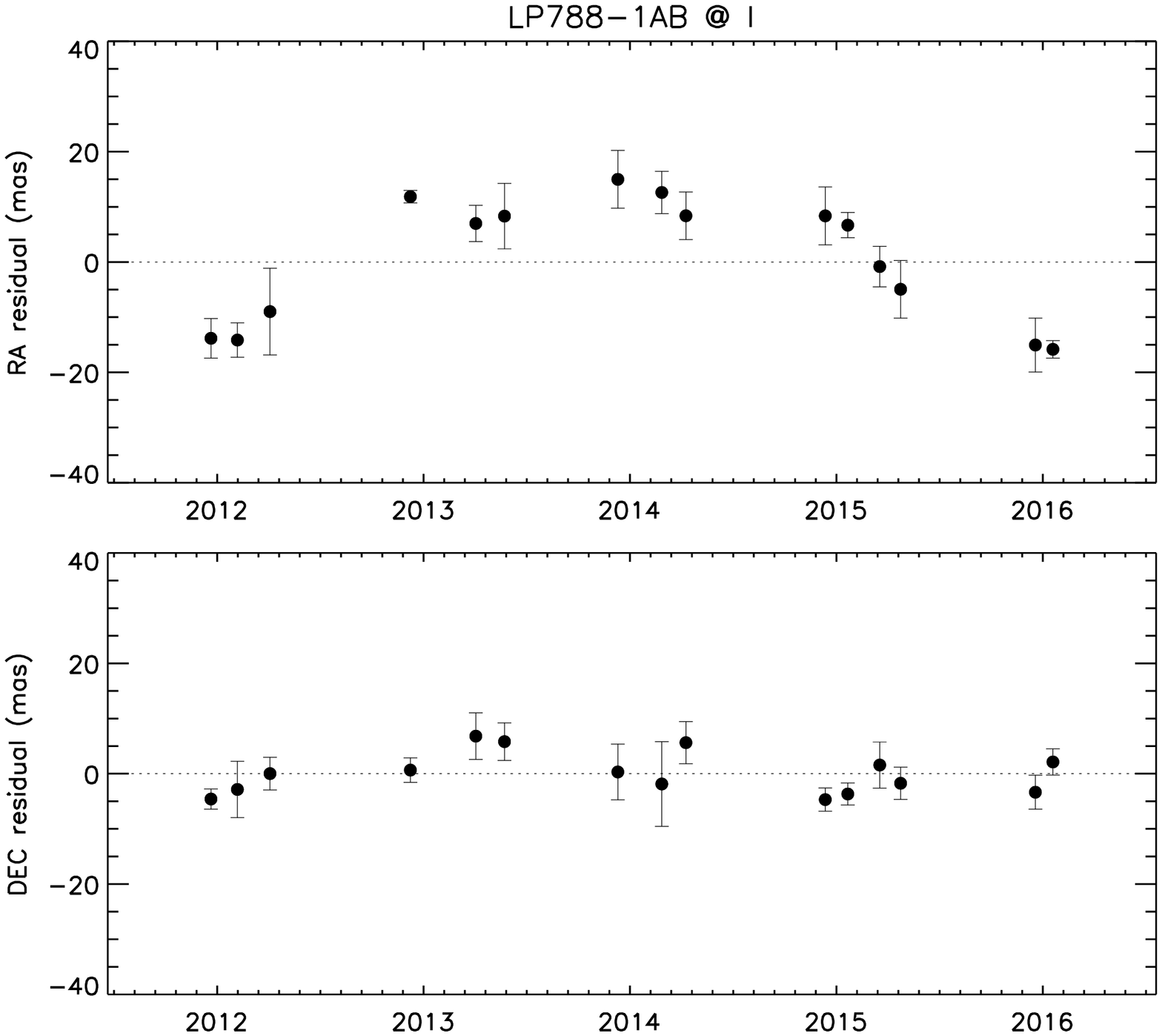}}
\endminipage\hfill
\minipage{0.50\textwidth}
\centering
{\includegraphics[scale=0.33,angle=90]{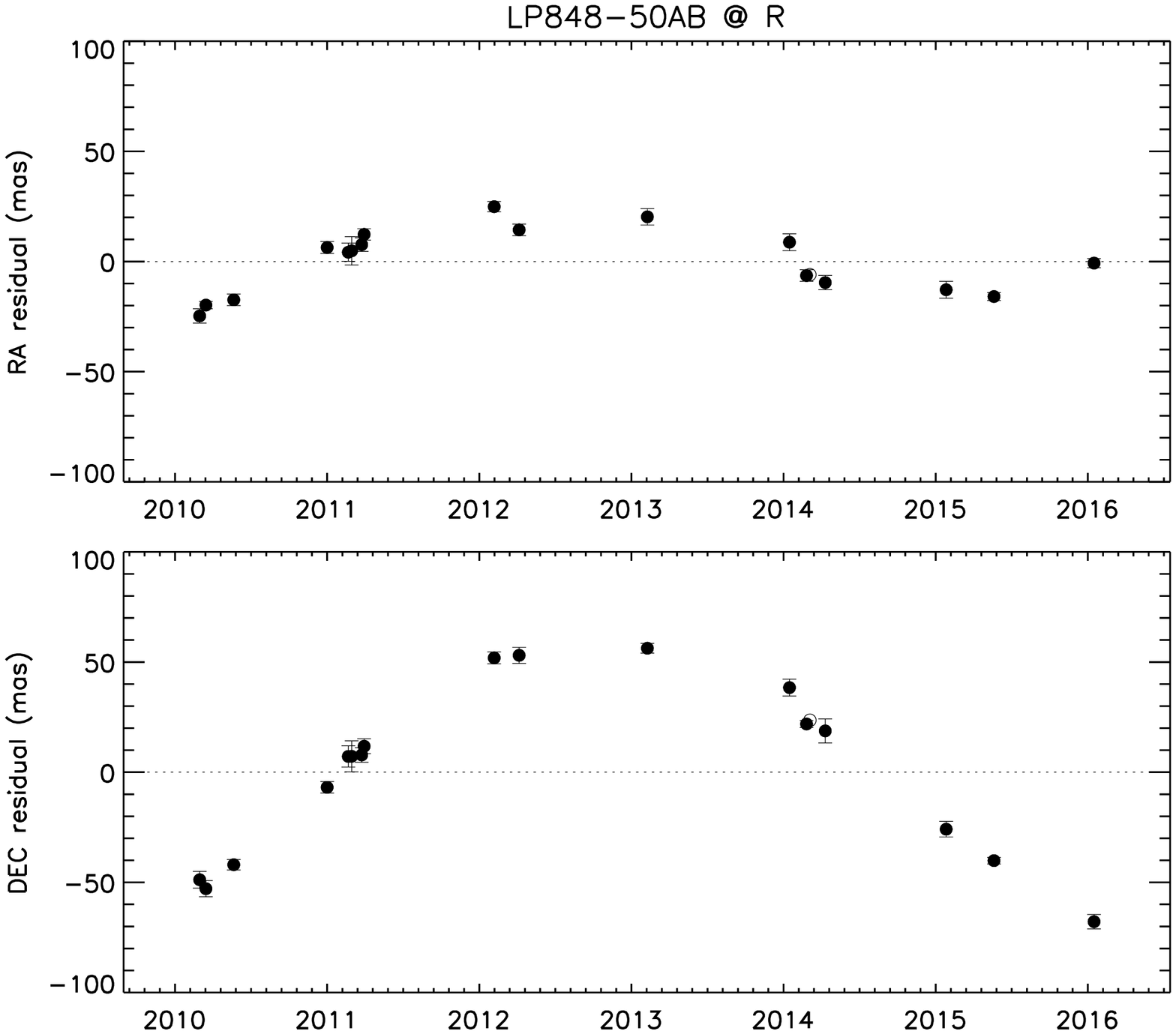}}
\endminipage\hfill
\vspace{15pt}
\minipage{0.50\textwidth}
\centering
{\includegraphics[scale=0.33,angle=90]{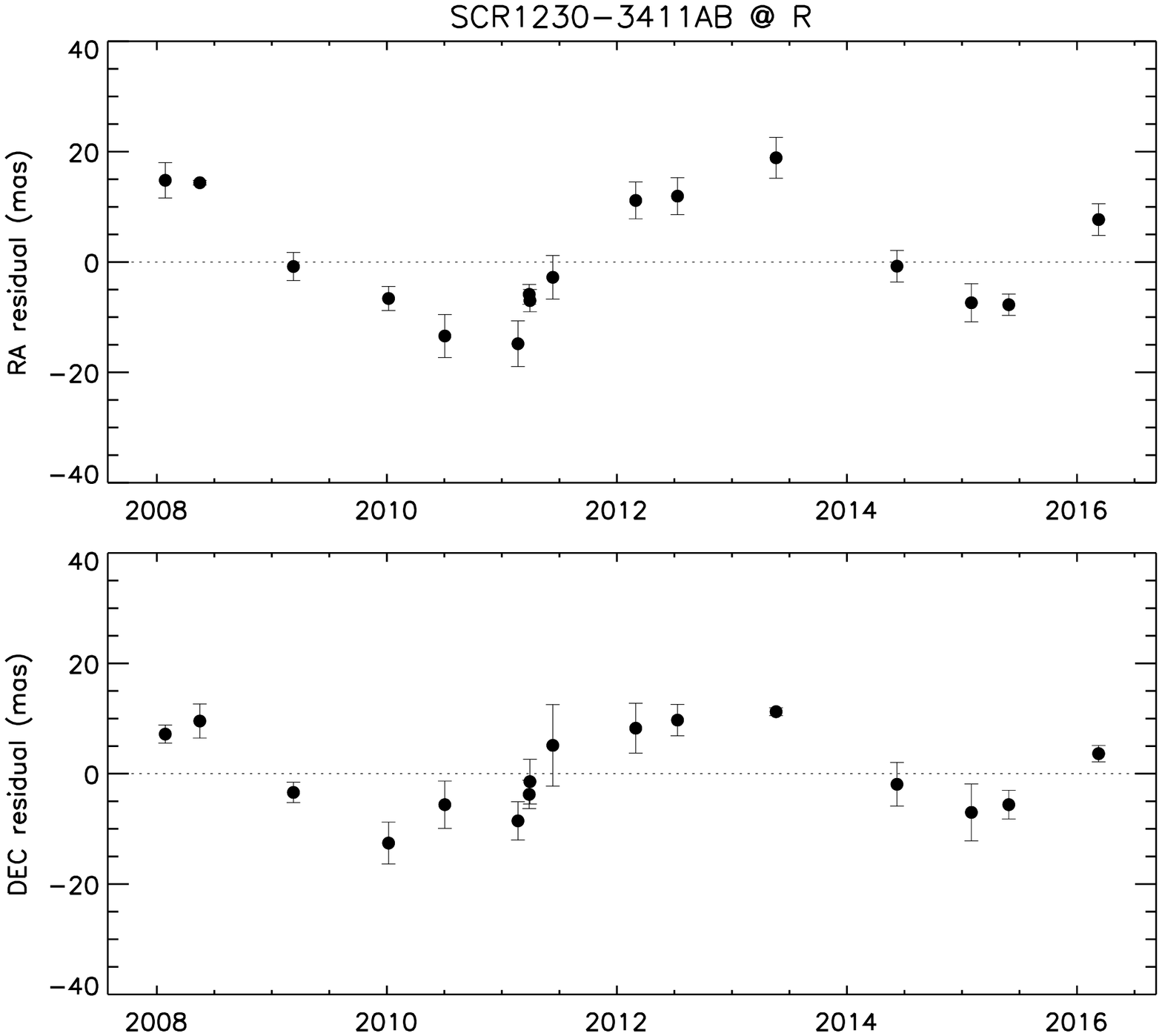}}
\endminipage\hfill
\minipage{0.50\textwidth}
\centering
{\includegraphics[scale=0.33,angle=90]{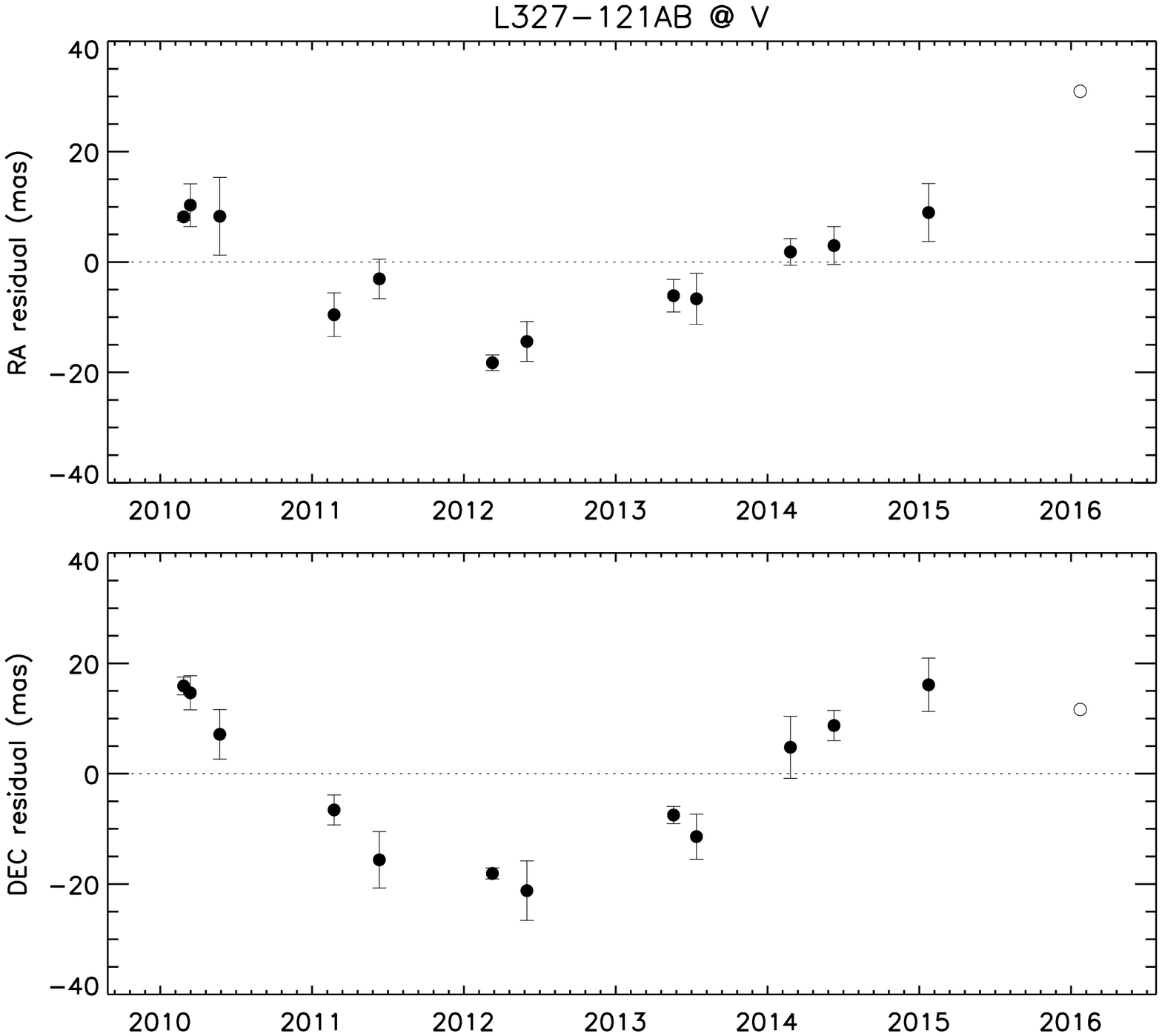}}
\endminipage\hfill
\vspace{15pt}
\minipage{0.50\textwidth}
\centering
{\includegraphics[scale=0.33,angle=90]{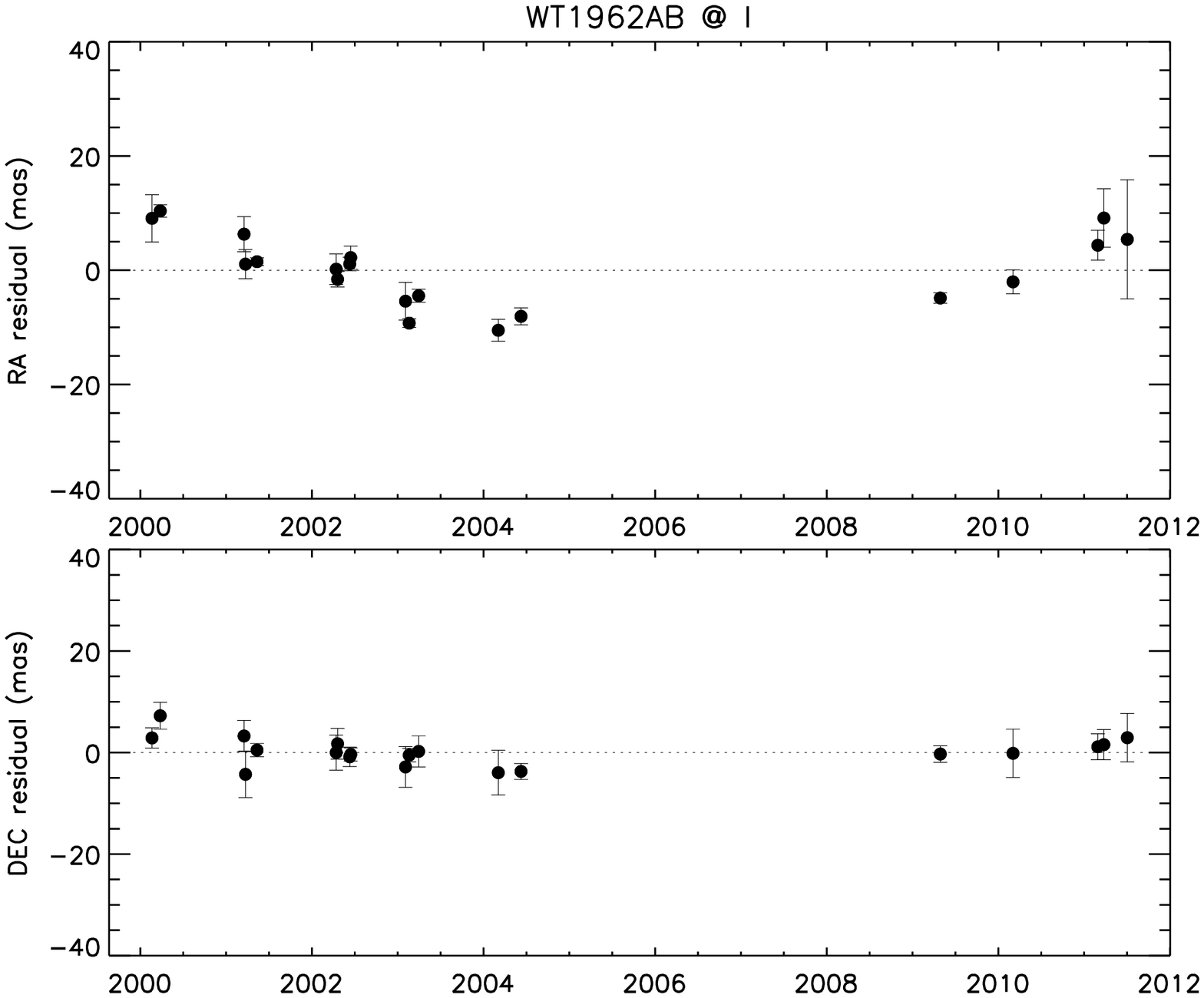}}
\endminipage\hfill
\minipage{0.50\textwidth}
\centering
{\includegraphics[scale=0.33,angle=90]{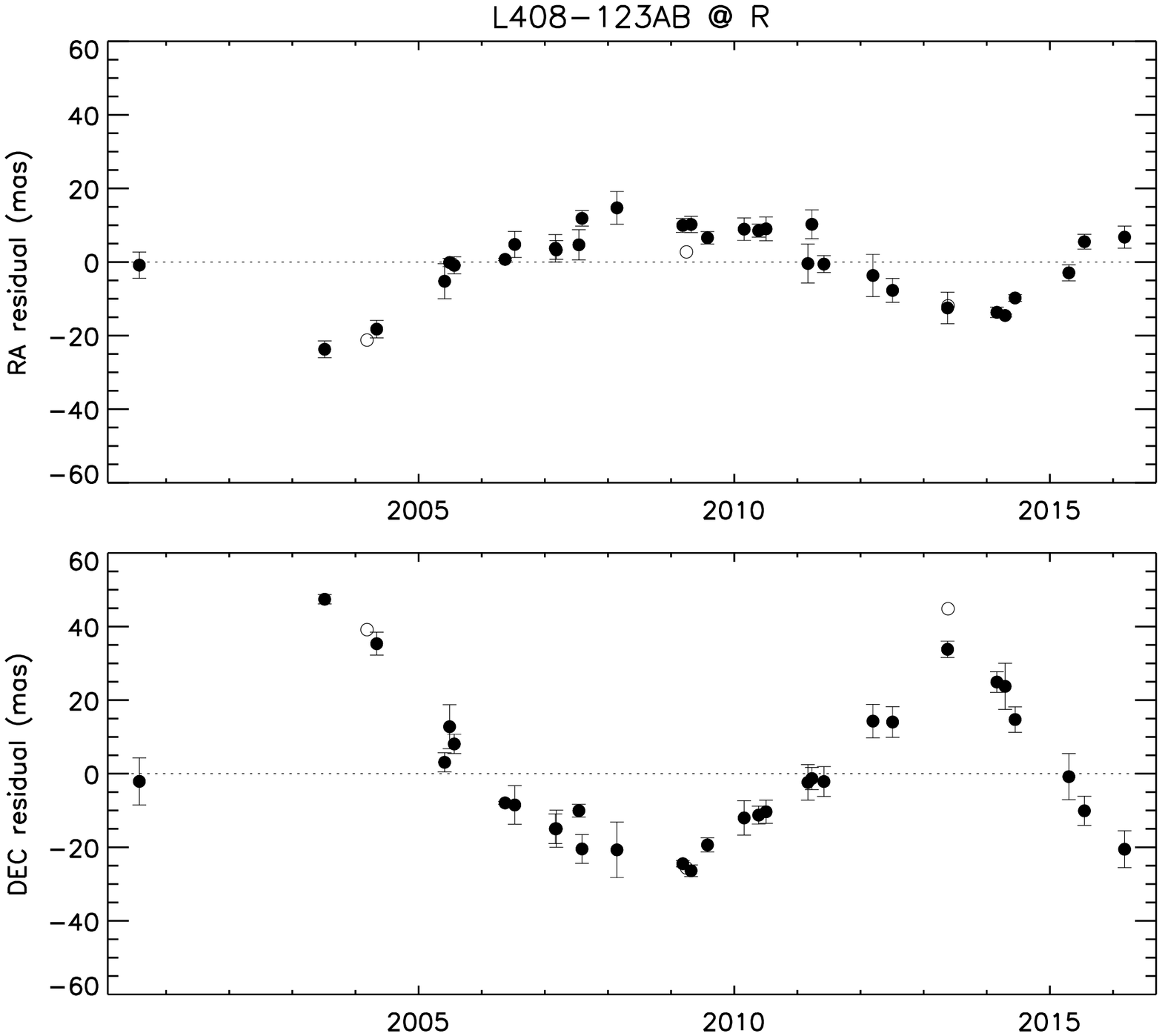}}
\endminipage\hfill
\vspace{10pt}
\caption[Perturbation Plots]{Nightly mean astrometric residual plots
  in RA and DEC for LP 788-1AB, LP 848-50AB, SCR 1230-3411AB, L
  327-121AB, WT 1962AB, and L 408-123AB. The astrometric signatures of
  each system's proper motion and parallax have been
  removed. \label{fig:perturbations2}}
\end{figure}

\begin{figure}[ht]
\minipage{0.50\textwidth}
\centering
{\includegraphics[scale=0.33,angle=90]{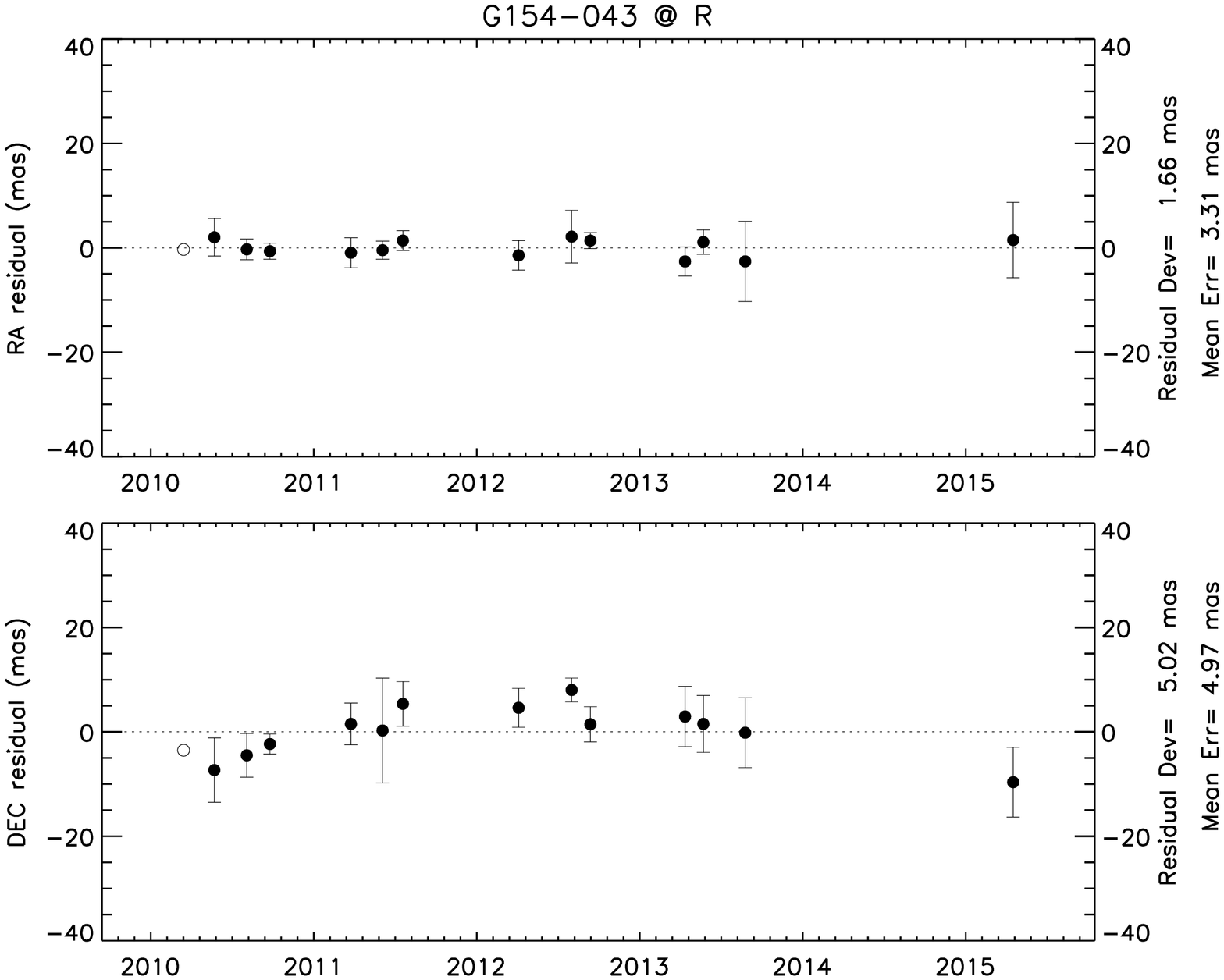}}
\endminipage\hfill
\minipage{0.50\textwidth}
\centering
{\includegraphics[scale=0.33,angle=90]{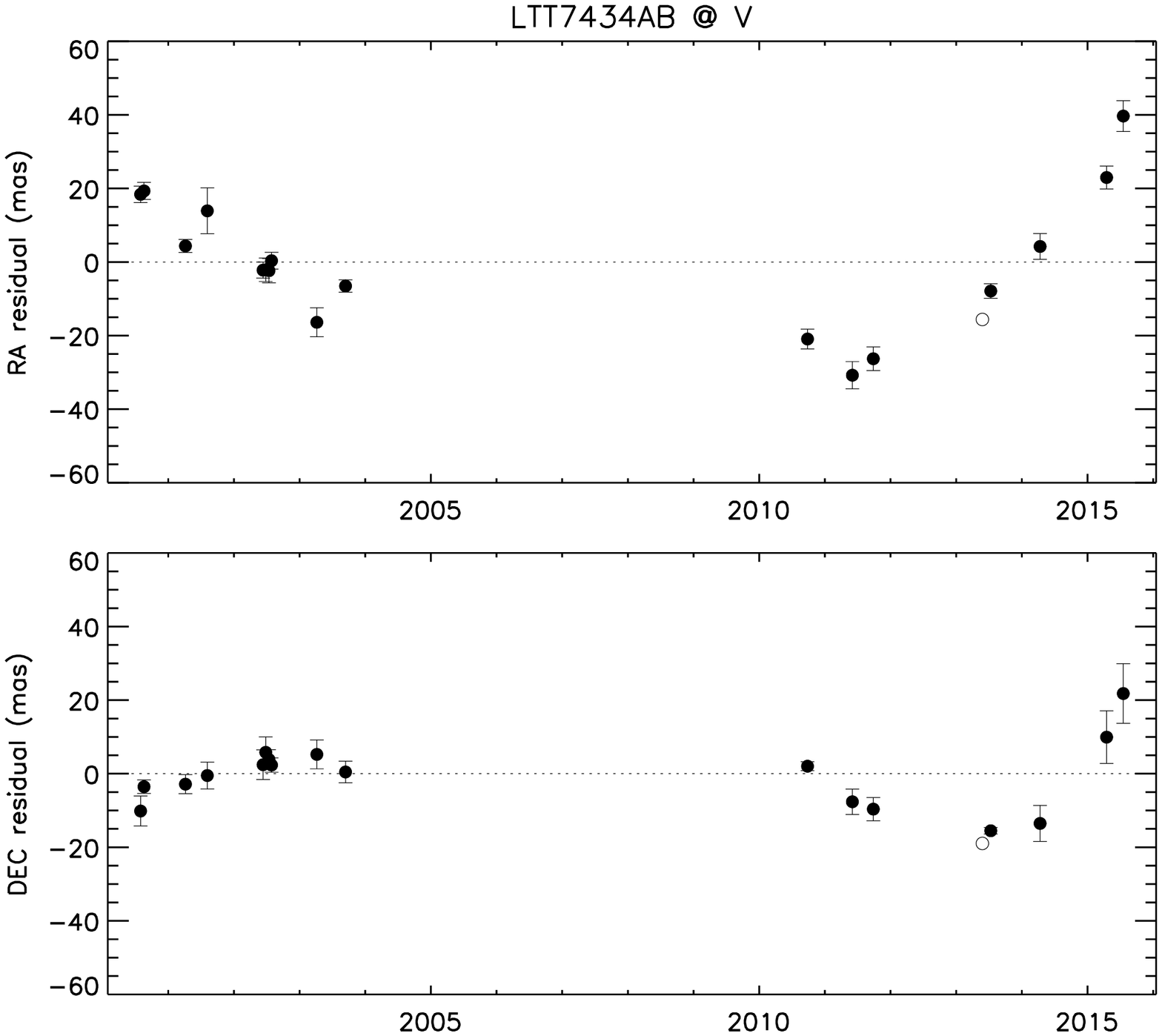}}
\endminipage\hfill
\vspace{15pt}
\minipage{0.50\textwidth}
\centering
{\includegraphics[scale=0.33,angle=90]{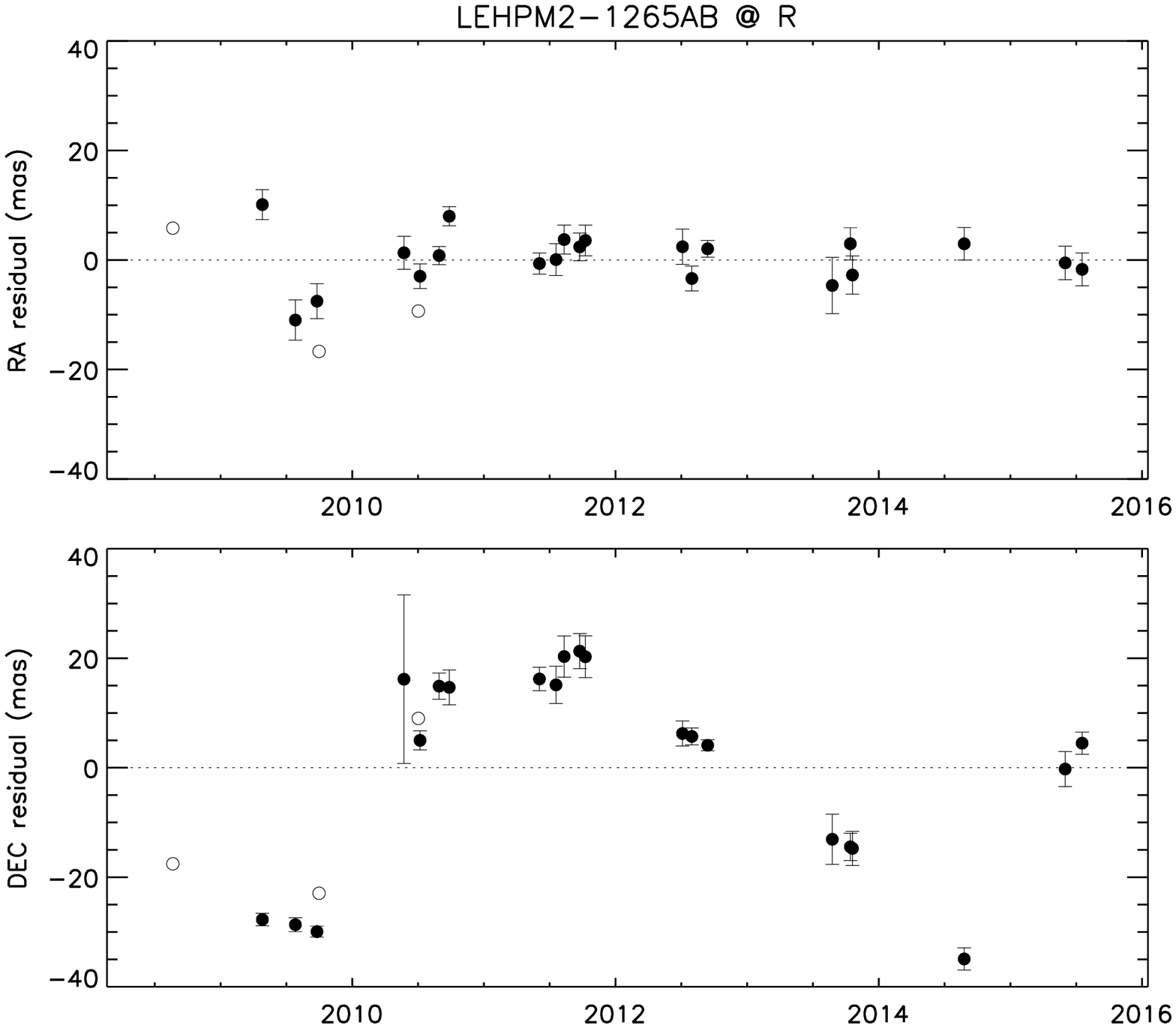}}
\endminipage\hfill
\minipage{0.50\textwidth}
\centering
{\includegraphics[scale=0.33,angle=90]{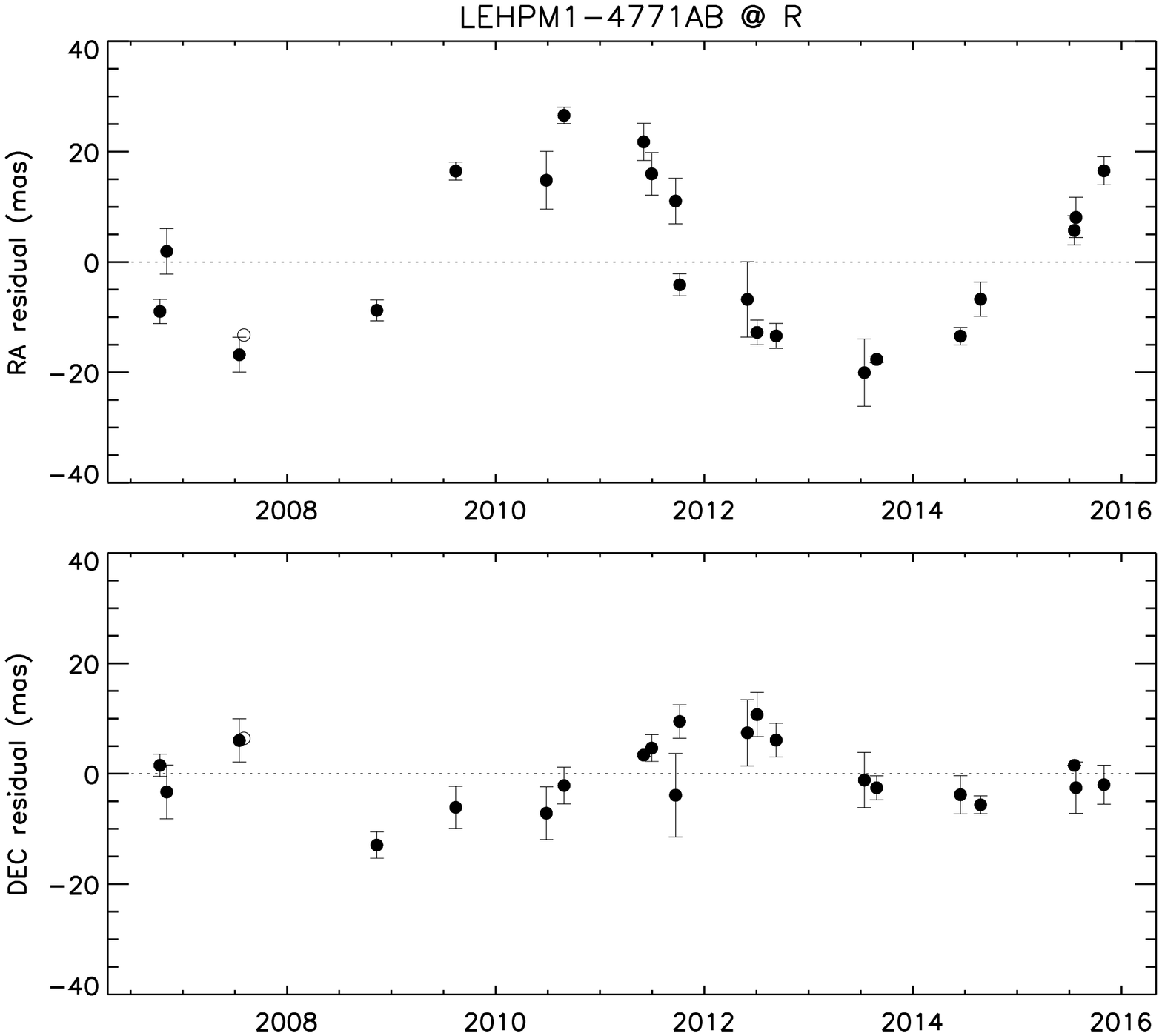}}
\endminipage\hfill
\vspace{15pt}
\minipage{0.50\textwidth}
\centering
{\includegraphics[scale=0.33,angle=90]{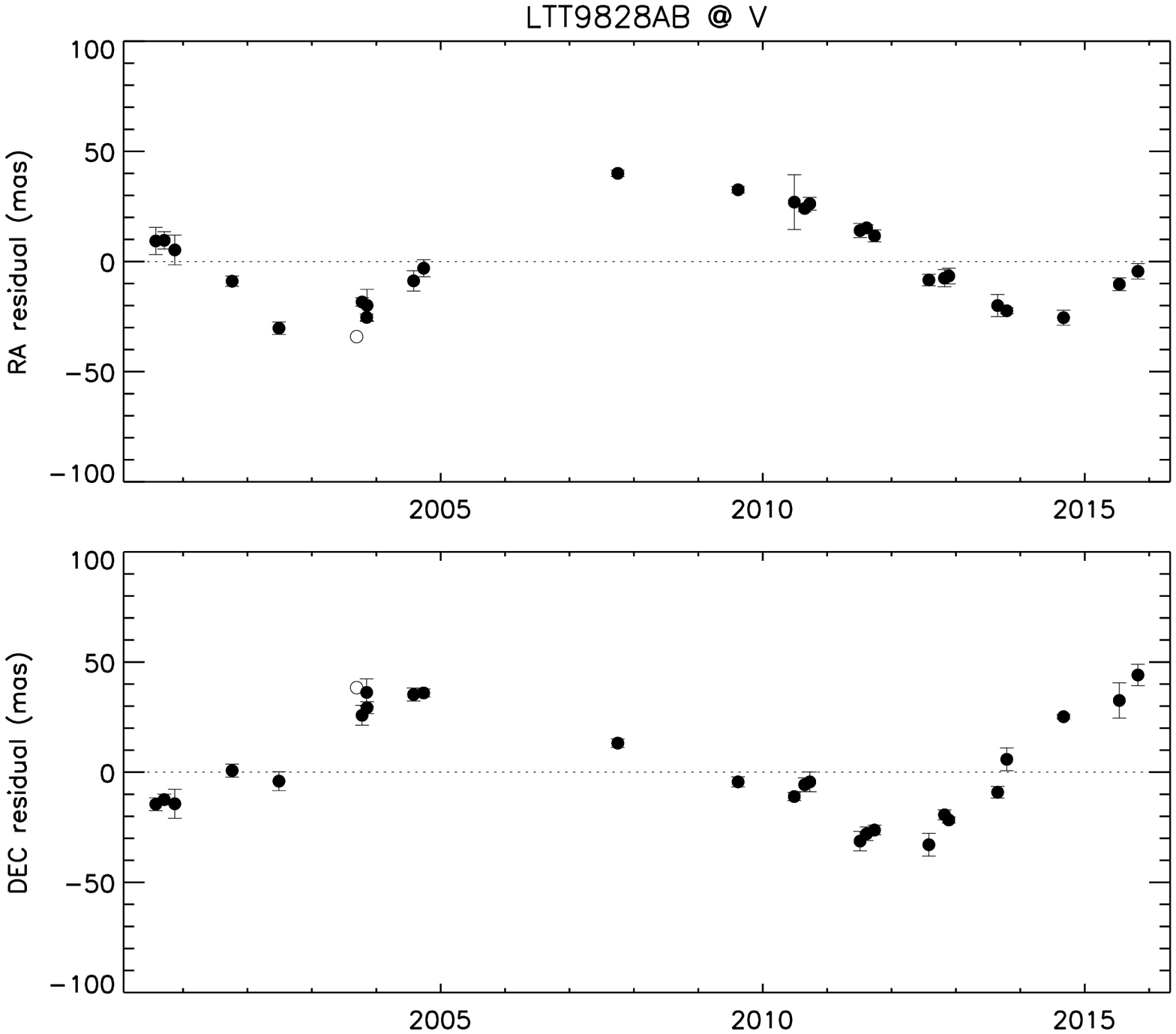}}
\endminipage\hfill
\vspace{10pt}
\caption[Perturbation Plots]{Nightly mean astrometric residual plots
  in RA and DEC for G 154-43AB, LTT 7434AB, LEHPM 2-1265AB, LEHPM
  1-4771AB, and LTT 9828AB. The astrometric signatures of each
  system's proper motion and parallax have been
  removed. \label{fig:perturbations3}}
\end{figure}


\begin{figure}[ht]
\minipage{0.50\textwidth}
\centering
{\includegraphics[scale=0.33,angle=90]{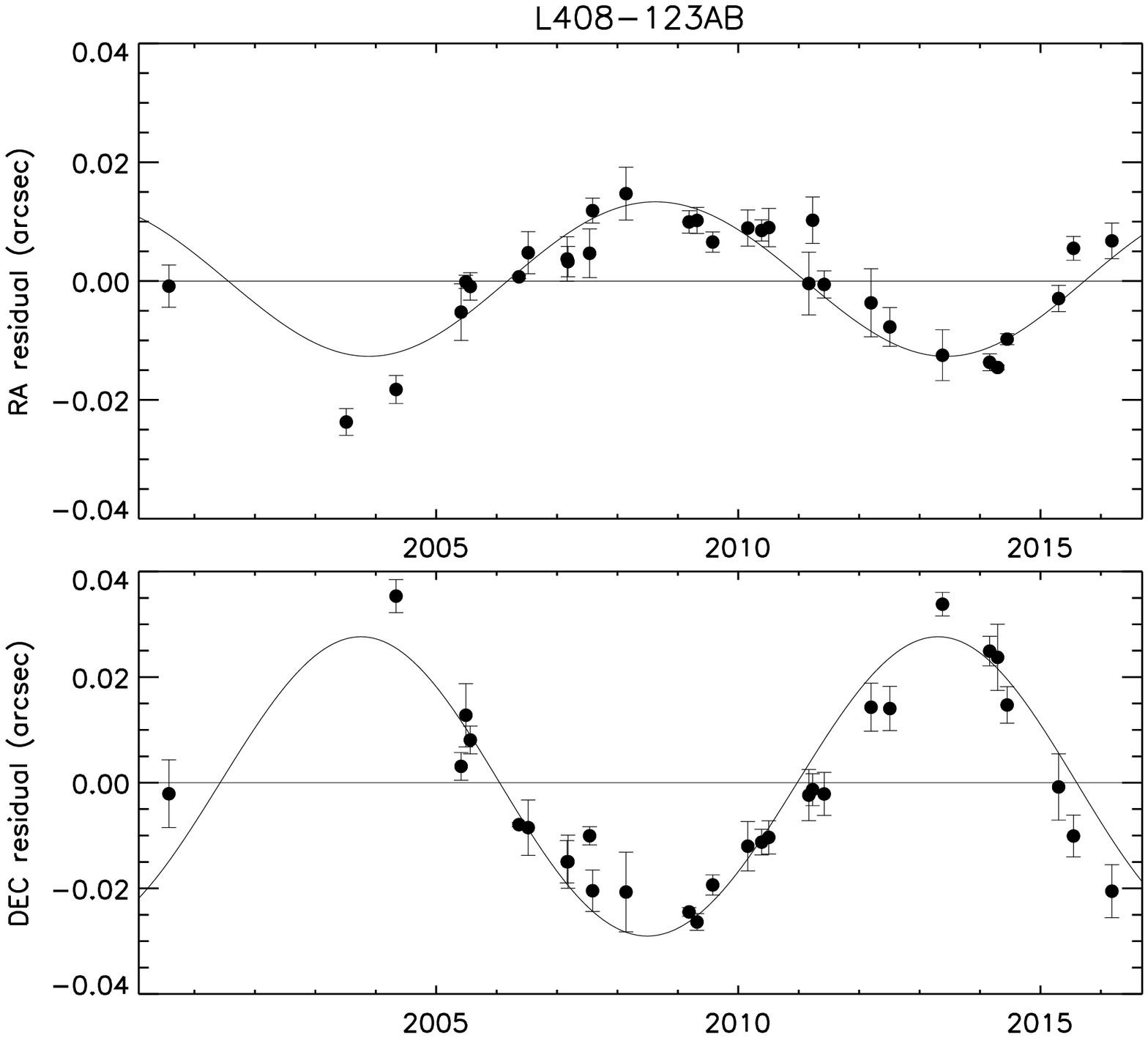}}
\endminipage\hfill
\minipage{0.50\textwidth}
\centering
{\includegraphics[scale=0.33,angle=90]{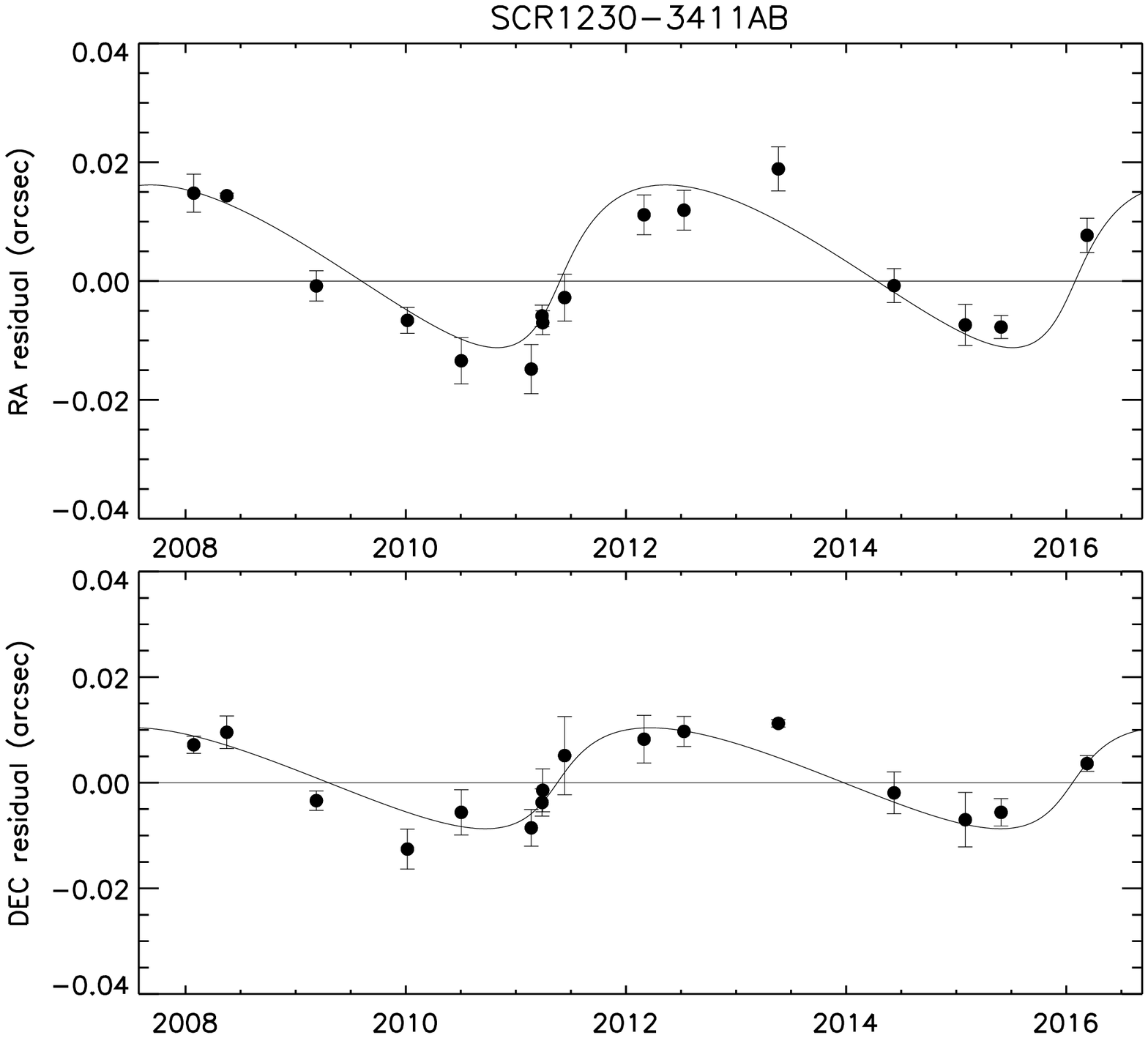}}
\endminipage\hfill
\minipage{0.50\textwidth}
\centering
{\includegraphics[scale=0.33,angle=90]{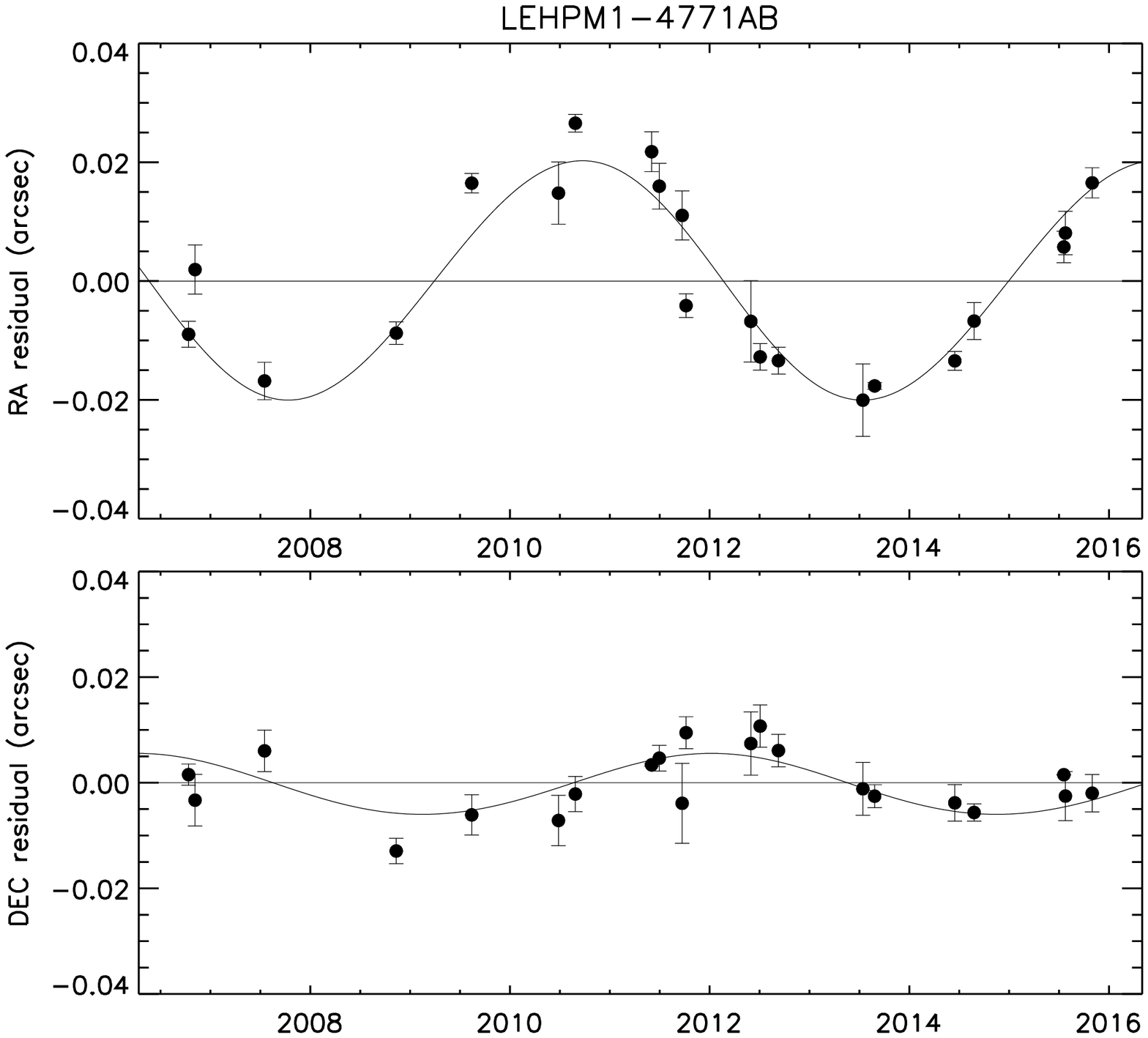}}
\endminipage\hfill
\minipage{0.50\textwidth}
\centering
{\includegraphics[scale=0.33,angle=90]{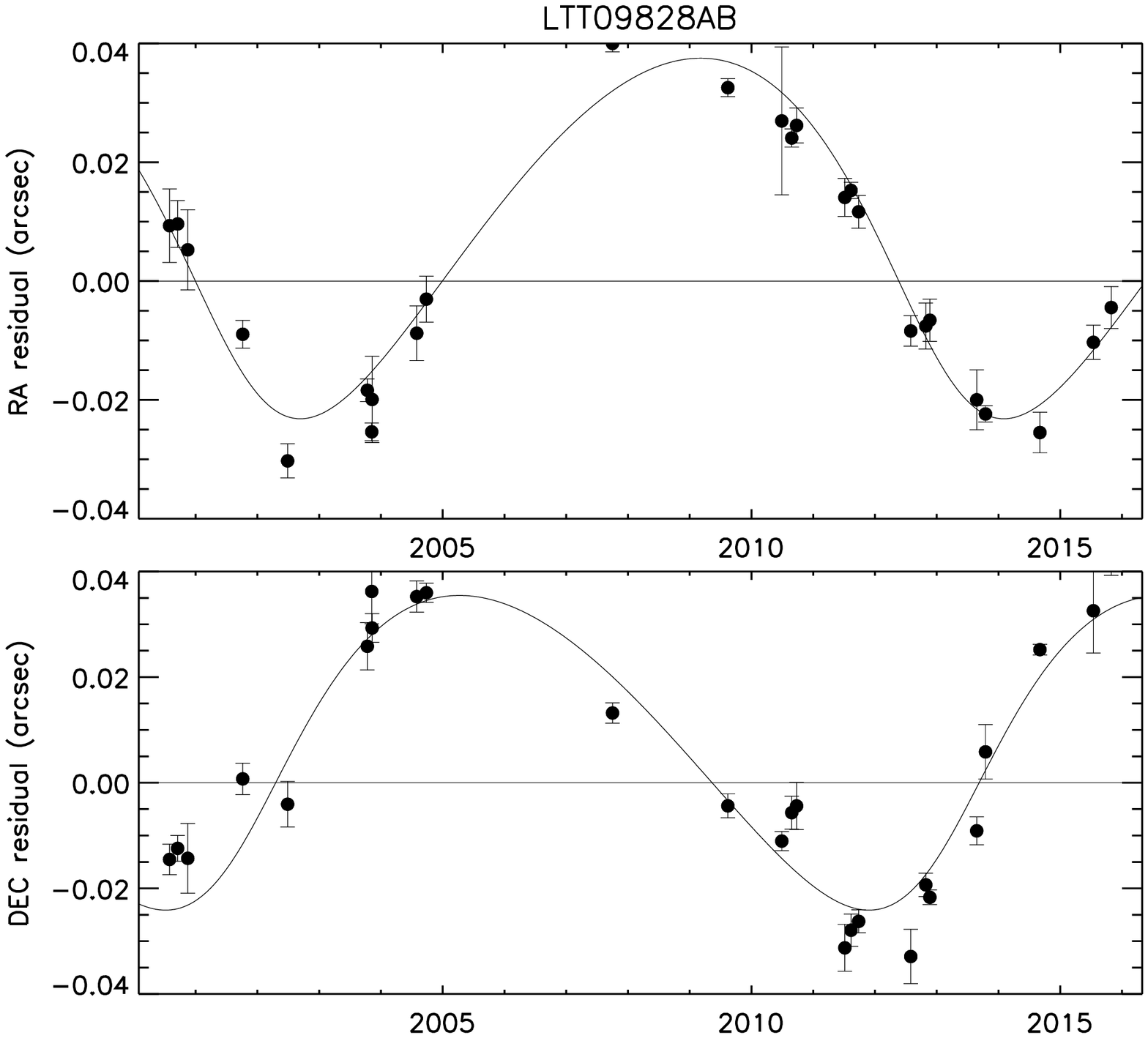}}
\endminipage\hfill
\caption{Orbital fits for four new binaries with complete
  orbits. Orbital elements are given in Table
  \ref{tab:orbits}.  \label{fig:orbits}}
\end{figure}


\begin{figure}
\includegraphics[scale=0.50,angle=90]{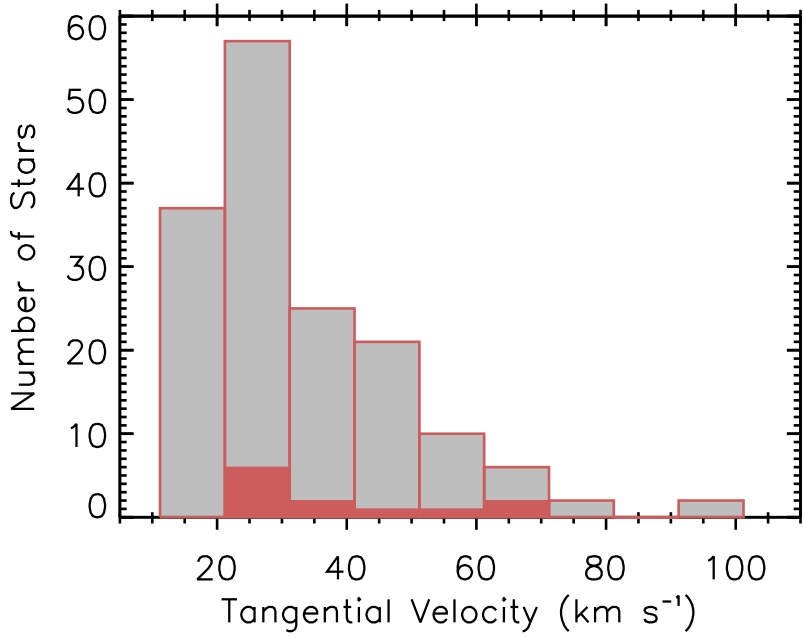}
\figcaption{Distribution of the tangential velocities of the the stars
  presented here. The 12 stars identified as candidate binaries are
  shown in pink. \label{fig:vtan}}
\end{figure} 


\begin{figure}
\includegraphics[scale=0.50,angle=90]{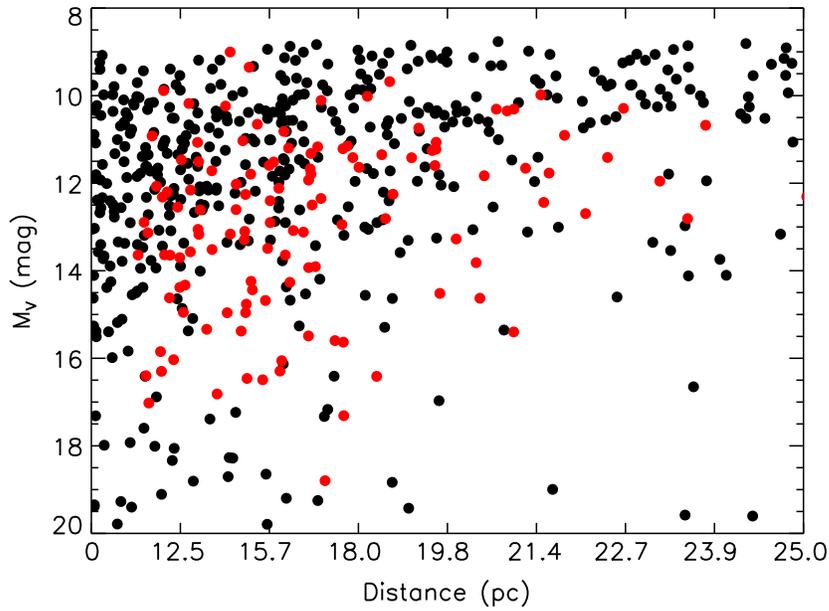}
\figcaption{Population density diagram of southern M dwarfs within 25
  pc, with distance to 25 pc indicated in intervals corresponding to
  eight equal-volume shells. Illustrated in red are the 123 new nearby
  (distances $<$ 25 pc) M dwarfs (in 116 systems) with accurate
  $\pi_{trig}$ from this work.  \label{fig:pop_den}}
\end{figure} 

\end{document}